\documentclass[letterpaper,twocolumn,10pt]{article}
\usepackage{usenix2019}

\usepackage{graphicx} 
\usepackage{xspace}
\usepackage{xcolor}
\usepackage{verbatim}
\usepackage{amsfonts}
\usepackage{booktabs}
\usepackage{multirow}
\usepackage[most]{tcolorbox}
\usepackage{subfigure}
\newcounter{insight}
\usepackage{courier}
\usepackage{xurl}
\usepackage[hidelinks]{hyperref}
\hypersetup{breaklinks=true}
\usepackage{url}
\usepackage[mathcal]{eucal}
\newcommand\sysname{\textit{NeuroImprint}\xspace}

\begin{document}

\title{From Efficiency to Leakage - \\ Privacy Backdoor in Federated Language Model Fine-Tuning}


\author{
{\rm Shanghao Shi$^{1}$, Chaoyu Zhang$^{2}$, Heng Jin$^{2}$, Yang Xiao$^{3}$,}\\
{\rm Yevgeniy Vorobeychik$^{1}$, William Yeoh$^{1}$, Ning Zhang$^{1}$, Y.\ Thomas Hou$^{2}$, Wenjing Lou$^{2}$}\\
$^{1}$Washington University in St.\ Louis \quad
$^{2}$Virginia Tech \quad
$^{3}$University of Kentucky
}

\maketitle

\begin{abstract}

Federated learning (FL) enables multiple parties to collaboratively fine-tune language models for domain-specific tasks without sharing raw data. Since full model fine-tuning is often prohibitively expensive for FL clients, parameter-efficient fine-tuning (PEFT) has become the de facto approach in practice, freezing the base model and training only a small set of adapters.
In this paper, we show that a malicious parameter server can stealthily corrupt a PEFT adapter into a privacy backdoor that implicitly memorizes the client’s training samples as \emph{isolated} per-sample parameter updates stored in \emph{separate neurons}, without degrading model utility. 
Concretely, our attack, \sysname, assigns a dedicated memorization neuron to each training sample and constrains that each neuron is updated at most once along the local fine-tuning trajectory. This design mitigates both cross-sample collisions and cross-step mixing introduced by large local batches and stateful optimizers (e.g., Adam/AdamW) in language-model fine-tuning. After fine-tuning, the resulting isolated per-sample updates can be analytically inverted in closed form to recover text embeddings, which are then deterministically mapped back to token sequences.
To understand the generality of our method, we implemented \sysname on multiple language models (BERT, GPT-2, Qwen2, and Llama3.2) and evaluated it across four fine-tuning datasets spanning diverse domains. The results demonstrate that our attack can reconstruct 59\% to 79\% of all finetuning samples with high semantic fidelity.

%
%
%
%
%
%

\end{abstract}


\section{Introduction}
Language models have demonstrated remarkable success across specialized domains such as healthcare, finance, legal assistance, and scientific research. Numerous domain-specific models have been proposed, including BioBERT \cite{lee2020biobert}, ClinicalBERT \cite{huang2019clinicalbert}, BioGPT \cite{luo2022biogpt}, and LLAVA-Med \cite{li2023llava} for healthcare applications; LegalBERT \cite{chalkidis2020legal} for legal tasks; FinBERT \cite{huang2023finbert}, FinGPT \cite{liu2023fingpt}, and FinLLAMA \cite{konstantinidis2024finllama} for financial analysis; as well as SciBERT \cite{beltagy2019scibert} for scientific research. Notably, these customized models are typically fine-tuned from large pre-trained foundation models on domain-specific datasets, rather than trained from scratch. In practice, this adaptation is mostly performed via parameter-efficient fine-tuning (PEFT) \cite{houlsby2019parameter, pfeiffer2020adapterhub, hu2023llm, he2021towards}, which freezes the backbone model and updates only lightweight modules (e.g., serial \cite{houlsby2019parameter}, parallel \cite{bapna2019simple, zhu2021counter}, or low-rank adapters \cite{hu2022lora, dettmers2023qlora}), reducing the computation, memory, and communication overhead of domain specialization. Since many domain-specific datasets contain sensitive or proprietary information (e.g., healthcare/financial records), and are strictly restricted from being shared beyond institutional borders due to regulatory constraints \cite{hipaa1996}.  Federated learning (FL) has been adopted as the de facto solution to enable distributed parties to collaboratively finetune language models without exposing the raw data \cite{mcmahan2017communication}. 
However, FL is known to be vulnerable to data reconstruction attacks, which aim to reconstruct the original training samples from model updates \cite{zhu2019deep, shi2023scale, zhao2024loki, fowl2021robbing, feng2024privacy}. This risk is increasingly concerning as language models are being deployed in safety- and privacy-critical applications.




\vspace{3pt} \noindent \textbf{Existing Literature and the Gap:}
Data reconstruction attack has been a well-studied topic in federated learning \cite{zhu2019deep, zhao2020idlg, geiping2020inverting, zhu2020r, yin2021see, lu2022april, hatamizadeh2022gradvit, pasquini2022eluding, wen2022fishing, fowl2021robbing, zhao2024loki, shi2023scale}. However, most existing attacks are developed for classical vision models and cannot handle discrete token sequences.
Recognizing the importance of this risk, recent efforts have begun to study data reconstruction attacks in language models \cite{carlini2021extracting, petrov2024dager}. 
Existing approaches either extract memorized training data from fully trained generative models via black-box querying or extend gradient inversion techniques to language models.
While these works demonstrate that language models can leak training data, they are largely insensitive to the fine-tuning stage or require access to raw gradients. 
More recently, prior work has introduced the concept of \emph{privacy backdoors}, showing that deliberate corruption of model initialization can induce data leakage during training \cite{wen2024privacy, feng2024privacy}. 
However, existing privacy backdoors are limited to membership inference or simple classifiers and do not extend to generative language models. 

More importantly, modern language model fine-tuning pipelines introduce several unaddressed challenges that fundamentally break existing approaches. 
First, fine-tuning is dominated by stateful optimizers such as Adam and AdamW, which entangle gradients across multiple steps and destroy the step-wise gradient information required for inversion. Second, fine-tuning operates over long discrete token sequences and large local batches under parameter-efficient fine-tuning (PEFT), requiring an attack design that can isolate and reconstruct many samples without interference.

\vspace{3pt} \noindent \textbf{Our Work:} 
We propose \sysname, a data reconstruction attack against federated fine-tuning of language models.
The fundamental idea is to stealthily craft a parallel PEFT adapter in the embedding layer as a privacy backdoor that “memorizes” per-sample parameter updates during fine-tuning, enabling closed-form reconstruction of local training samples afterward. Our attack addresses the following challenges:

\vspace{3pt} \noindent \textit{Challenge 1---Reconstructing long discrete token sequences}. Unlike vision, language inversion is challenging because the reconstruction targets are long \emph{discrete} token sequences, where small token-level errors can break syntax and semantics. Our guiding principle is to avoid direct optimization in the discrete token space and instead operate in the continuous embedding space whenever possible. Accordingly, we transform an unstable discrete search into a principled two-stage procedure: exact (or near-exact) analytic recovery in the embedding space followed by a deterministic mapping back to text.

\vspace{3pt} \noindent \textit{Challenge 2---Inversion under stateful optimizers.}
Stateful optimizers such as Adam/AdamW fundamentally complicate inversion because their momentum and adaptive-variance states accumulate information across many local steps, entangling per-sample gradients over time. Our guiding principle is that the optimizer state becomes irrelevant if each parameter is updated only once. We therefore design the backdoor so that each memorization neuron is activated, and hence updated, by at most one training sample over the entire local training trajectory. This enforces \emph{temporal single-sample activation}, eliminating cross-step mixing and turning inversion from a hard multi-step problem into a tractable single-step reversion.

\vspace{3pt} \noindent \textit{Challenge 3---Scaling reconstruction to large fine-tuning batches.}
Scaling reconstruction is difficult because when many samples contribute updates in the same round, their information collides unless there is an explicit separation mechanism. Our principle here is to allocate independent memorization capacity per sample. Concretely, we partition the backdoor into many independent reconstruction slots such that each sample is routed to a distinct memorization neuron, while all token-level information of that sample is captured within that neuron. This \emph{one-neuron--one-sample} organization prevents cross-sample interference and enables large-batch reconstruction, allowing an effective trade-off between space and attack scalability.

\vspace{3pt} \noindent \textit{Challenge 4---Stealthy backdoor without utility loss}. A practical attack must keep fine-tuning behavior and model performance unchanged to avoid detection. Our guiding principle is to leverage architectural invariances in the host model, in particular the normalization invariance of LayerNorm. \textit{\sysname} achieves this by crafting its output layer with identical row vectors and fixed biases, so that the subsequent LayerNorm provably cancels the backdoor’s contribution through deterministic normalization.

We implemented \textit{\sysname} on four different language models, including BERT \cite{devlin2019bert}, GPT-2 \cite{radford2019language}, Qwen 2-1.5B \cite{bai2023qwen}, and Llama 3.2-3B \cite{dubey2024llama}, and on four datasets, including AGNews \cite{zhang2015character}, SQuAD \cite{rajpurkar2016squad}, EMRQA-mSQuAD \cite{eladio2024emrqa}, and GSM8K \cite{cobbe2021gsm8k}. The results demonstrate that \textit{\sysname} can recover 59\% to 79\% of the finetuning samples under general settings with high semantic similarity. We further explore the system and attack factors that may affect the reconstruction performance. Particularly, we validate that the model utility will not be affected by the backdoor component, demonstrate the decent cross-dataset attack transferability under various settings, and illustrate the effectiveness of \textit{\sysname} under LoRA finetuning, even when applied to larger models such as Qwen2-7B and Llama3.2-8B. 

\vspace{3pt} \noindent \textbf{Contributions:} 
\begin{itemize}
    \item We propose \textit{\sysname}, a novel data reconstruction attack against federated language model fine-tuning. \textit{\sysname} overcomes the unique challenges of federated fine-tuning and undermines the privacy-preserving property of the FL paradigm.
    \item We provide rigorous mathematical analysis and attack insights for \textit{\sysname}, which highlight its distinctive properties and deepen the understanding of FL privacy attacks. Notably, \textit{\sysname} is a \textit{closed-form} attack backed by formal reconstruction guarantees. 
    \item We implement \textit{\sysname} on four language models and evaluate it across four datasets across diverse attack domains. The results confirm the \textit{effectiveness} and \textit{practicality} of \sysname in real-world fine-tuning scenarios. 
\end{itemize}

\section{Background}

\subsection{Language Modeling}

Language models are the backbones of modern natural language processing. They function as auto-regressive models to predict the next token $x_t$ given the input sequence $(x_1, x_t, \cdots, x_{t-1})$. A language model usually consists of an embedding layer $E$ followed by a stack of transformer blocks $(T_1, T_2, \cdots, T_{l})$. The embedding layer maps the distributed token sequence $s$ into continuous embedding vectors $e \in \mathbb{R}^{s\times m}$, where $m$ refers to the hidden dimension. After this, the transformer blocks process the embedding vector sequentially to generate the desired output.   

\begin{figure}[t]
    \centering
    \includegraphics[width=0.45\textwidth]{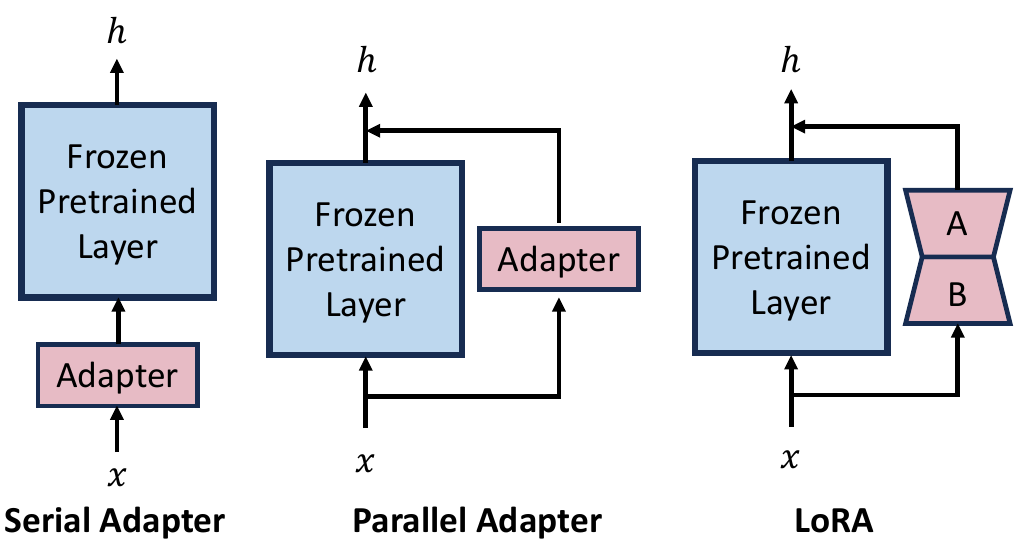}
    \caption{
    Serial, parallel, and low-rank adapters. }
    \label{fig: PEFT-adapters} 
\end{figure}

\textbf{Parameter Efficient Fine Tuning (PEFT):} Current language models are becoming larger and larger, making it prohibitively expensive to train from scratch. As a result, adapting a pretrained model to downstream tasks is typically done by fine-tuning on small, task-specific datasets, which are often proprietary or sensitive and thus raise privacy concerns if exposed. In practice, this adaptation is commonly performed via parameter-efficient fine-tuning (PEFT) \cite{houlsby2019parameter, pfeiffer2020adapterhub, hu2023llm, he2021towards}, which freezes the backbone model and updates only lightweight modules to reduce unnecessary computation, memory, and communication overhead. Representative PEFT designs include \emph{serial adapters} inserted between frozen sublayers \cite{houlsby2019parameter}, \emph{parallel adapters} added as parallel branches \cite{bapna2019simple, zhu2021counter}, and \emph{low-rank adapters} such as LoRA/QLoRA that parameterize weight updates with low-rank factors \cite{hu2022lora, dettmers2023qlora}. We demonstrate them in Fig. \ref{fig: PEFT-adapters}.


\textbf{Federated Learning:} To avoid exposing raw data, federated learning (FL) enables a parameter server $S$ and a set of clients $c_k$ to collaboratively finetune language models, where $k\in\{1,2,\cdots,n\}$. In each round $r$, the parameter server $S$ first publishes the global model parameter $\theta^{r}$ to a subset of selected clients $\mathcal{C}^{r}\subseteq \mathcal{C}$. 
Then these clients compute the local model updates $\theta_i^r$ with their local datasets $D_i$ under multiple local training steps $T$. When using the SGD optimizer, $\theta_i^r$ can be computed iteratively as follow: $\theta_i^{r,t}=\theta_i^{r, t-1}-\eta \nabla \mathcal{L}(\theta_i^{r, t-1}, D_i)$, where $\mathcal{L}$ refers to the loss function, $\eta$ refers to the learning rate, and $t\in\{1,2,\cdots, T\}$. When using the Adam/AdamW optimizer, $\theta_i^r$ is computed as $\theta_i^{r,t} = \theta_i^{r, t-1} - \eta \frac{\hat{m}_i^t}{\sqrt{\hat{v}_i^t} + \varepsilon}$, where $\hat{m}_i^t$ and $\hat{v}_i^t$ refers to the intermediate training states. After this, the parameter server aggregates all local updates with the FedAVG algorithm \cite{mcmahan2017communication} to proceed with the training process:
\begin{equation}
    \begin{aligned}
        \theta^{r+1}=\frac{1}{n}\sum_{i=1}^n \theta_i^r
    \end{aligned}
\end{equation}
More detailed formulations can be found in Appendix A. 

\subsection{Data Reconstruction Attacks \& Defenses}

Data reconstruction attacks focus on extracting training samples from machine learning models. Under this big umbrella, several specialized variants have been developed. 

\textit{Optimization-based inversion attacks} assume an honest-but-curious parameter server and aim to reconstruct training data by reversing raw gradients through iterative optimization \cite{zhu2019deep, zhao2020idlg, geiping2020inverting, zhu2020r, yin2021see, lu2022april, hatamizadeh2022gradvit}. These methods typically initialize random dummy inputs and gradually update them by aligning their gradients with the observed ones, driving the dummies toward the original training samples. However, their reconstruction capability is limited to a dozen and can be easily defended by the secure aggregation mechanisms.

\textit{Model crafting-based inversion attacks} assume stronger threat models in which the parameter server actively modifies model parameters or architectures to improve the attack scalability and bypass secure aggregation mechanisms \cite{pasquini2022eluding, wen2022fishing, fowl2021robbing, zhao2024loki, shi2023scale, petrov2024dager, feng2024privacy}. State-of-the-art works \cite{zhao2024loki, shi2023scale} have demonstrated the capability of reconstructing hundreds of training samples with high fidelity. However, these works are confined to vision models and can only reverse SGD-style model updates, which are not mainstream for language model finetuning.  

\textit{Generative inversion attacks} aim to re-generate training samples from a trained model using extraction prompts \cite{peng2024pseudo, zhang2020secret, carlini2023extracting, carlini2021extracting}. Prior work shows they can recover thousands of high-fidelity training examples from commercial diffusion models \cite{carlini2023extracting} and large language models \cite{carlini2021extracting}. However, such attacks reconstruct samples from the entire training set without guarantees for any specific example, making them unsuitable to target the fine-tuning set, which represents only a small fraction of the data



\textit{Prompt inversion attacks} target the collaborative inference system and aim to reverse the shared hidden states between different entities back to user input prompts \cite{he2019model, qu2025prompt, dong2025depth}. These works demonstrate strong performance in inverting deep hidden states back to input prompts. However, intermediate hidden states are not available in the FL training process, rendering these works ineffective. 

\textit{Privacy backdoors} stealthily corrupt model parameters to induce private data leakage \cite{wen2024privacy, feng2024privacy}. Compared to conventional security backdoors that aim to trigger misclassifications or detection failures \cite{chen2017targeted, gu2017badnets, liu2018trojaning, carlin2022Poison}, recent work implants privacy backdoors through supply-chain attacks by corrupting pre-trained models before downstream fine-tuning. However, existing methods are limited to membership inference \cite{wen2024privacy} or simple classifiers and do not extend to text-generation models \cite{feng2024privacy}.

\begin{figure}[t]
    \centering
    \includegraphics[width=0.47\textwidth]{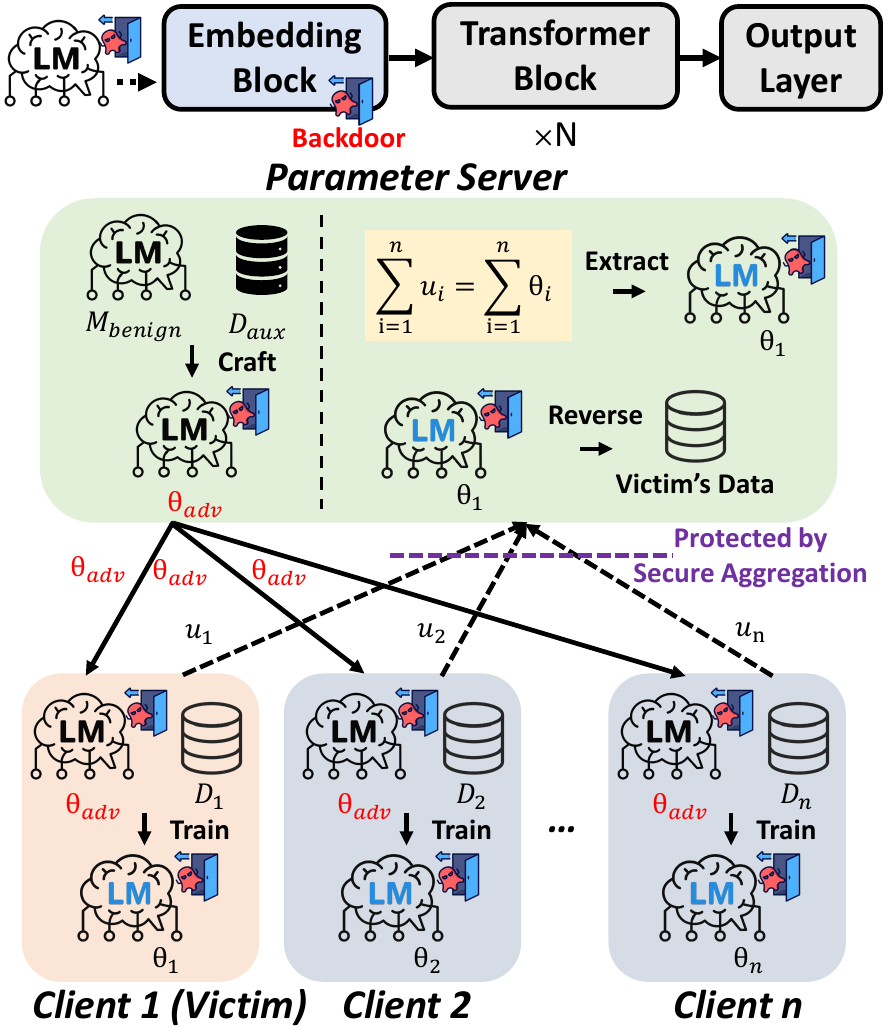}
    \caption{\sysname threat model. We remove the superscript $r$ that represents the FL round number because \sysname can be accomplished within one round.
    }
    \label{fig: threat-model} 
\end{figure}

\textit{Secure aggregation} is considered the state-of-the-art defense mechanism for protecting federated learning systems \cite{bonawitz2017practical}, whose fundamental technique is to add zero-sum cryptographic masks to individual clients' model updates. As a result, the curious server can only obtain the aggregated model updates and cannot infer individual information. Beyond vanilla secure aggregation mechanisms, there is a series of literature focusing on improving the communication and computation overheads \cite{bell2020secure, guo2020v, kadhe2020fastsecagg}, and enhancing the resilience against malicious attacks \cite{burkhalter2021rofl, rathee2023elsa, bell2023acorn, ma2023flamingo}. However, existing attacks have demonstrated the capability to bypass the secure aggregation mechanism, and we also take this as one of our attack goals.

\section{Threat Model}

\textbf{System Model:} We consider a federated fine-tuning system with a parameter server $S$ and multiple clients $c_i$, where each client holds a local dataset $D_i$. To reduce the client-side training overhead, the system applies PEFT by freezing the base model and training lightweight adapters within selected blocks. We further assume the secure aggregation is deployed to hide individual client updates $\theta_i$.

\noindent\textbf{Adversary's Objectives:} The adversary is the parameter server $S$. Its goal is to bypass secure aggregation and reconstruct a substantial portion of a target client’s fine-tuning dataset $D_{\mathrm{target}}$. To remain stealthy, the adversary implants a privacy backdoor by corrupting only the embedding-layer PEFT adapter $\Delta_{\mathrm{adv}}$, leaving all other transformer blocks and their adapters unchanged. The adversary also aims to cause no or negligible performance degradation.


\noindent\textbf{Adversary's Capabilities:} We consider an adversary that can maliciously initialize the linear layer and non-linear activation functions of the embedding-layer PEFT adapter, turning it into a privacy backdoor. In particular, we assume this PEFT module is implemented as a \emph{parallel adapter} attached to the embedding block. After fine-tuning, since the system is protected by secure aggregation, we assume that the adversary can only access the aggregated model update $\sum_{i=1}^n u_i=\sum_{i=1}^n \theta_i$. 
We assume the adversary can exploit an auxiliary dataset $D_{\mathrm{aux}}$, collected from publicly available sources, to help craft the configuration of $\Delta_{\mathrm{adv}}$. We follow the common setting of existing attacks \cite{zhao2024loki, shi2023scale, fowl2021robbing, feng2024privacy}, that this auxiliary dataset $D_{\mathrm{aux}}$ has a similar distribution as the fine-tuning dataset $D_{\mathrm{target}}$. However, we acknowledge that in practice, the adversary may only have generic knowledge of the target dataset, and there can be distribution shifts between the two datasets. We comprehensively investigate the impact of such distribution shifts in Section \ref{sec: evaluation}. We demonstrate the detailed threat model in Fig. \ref{fig: threat-model}. 

\section{Attack Methodology}
\label{sec: method}

\subsection{Attack Warm-up}
To better illustrate our approach, we start by introducing the closed-form inversion against a simple linear layer \cite{fowl2021robbing, shi2023scale}:
\begin{equation}
        \mathbf{h}=\mathrm{ReLU}(\mathbf{W}\mathbf{x}+\mathbf{b})
\end{equation}
where $\mathbf{x}\in \mathbb{R}^s$ is the layer's input, $\mathbf{W}\in \mathbb{R}^{m\times s}$ is the weight matrix, $\mathbf{b}\in \mathbb{R}^m$ is the bias vector, and $\mathbf{h}\in\mathbb{R}^m$ is the output.

When the training loss $\mathcal{L}$ is backpropagated to the linear layer and the $i^{th}$ neuron is activated, the gradients of the corresponding $i^{th}$ row of the weight matrix, denoted by $\mathbf{w}_i$,  and the $i^{th}$ entry of the bias vector, denoted by $b_i$, can be derived as follows: 
\begin{equation}
    \nabla_{\mathbf{w}_i}\mathcal{L}=\frac{\partial \mathcal{L}}{\partial h_i}\cdot \mathbf{x}, \quad \nabla_{b_i}\mathcal{L}=\frac{\partial \mathcal{L}}{\partial h_i} 
\end{equation}
Take a step further, the input $\mathbf{x}$ can be analytically reconstructed by the scalar division of the two gradients. 
\begin{equation}
\label{equ: gradient-invert}
\mathbf{x}=\nabla_{\mathbf{w}_i}\mathcal{L}/\nabla_{b_i}\mathcal{L}
\end{equation}
Such a reconstruction is not an approximation of the inputs but an \textit{exact} analytical recovery. Based on this, we have the following attack insight:

\begin{tcolorbox}[colback=gray!10, colframe=gray!50]
\refstepcounter{insight}\label{insight: gradient-invert}
\textbf{Insight 1}: The gradients of an activated neuron in a linear layer can explicitly \textit{memorize} and \textit{recover} a specific training sample.
\end{tcolorbox}
However, Eq. \ref{equ: gradient-invert} applies only to a single input sample and fails for batched or aggregated inputs, as their gradients are mixed during backpropagation, causing reconstruction conflicts.
To overcome this limitation, prior works \cite{fowl2021robbing, zhao2024loki, shi2023scale} leverage an auxiliary dataset $D_{\mathrm{aux}}$ that shares the same data distribution as the target dataset $D_{\mathrm{target}}$. Since Eq. \ref{equ: gradient-invert} only utilizes a single neuron from the linear layer, these works keep the row vector $\mathbf{w}_i$ fixed while varying the bias term $b_i$
across $D_{\mathrm{aux}}$ to create $m$ reconstruction neurons that collectively span the full distribution ($i \in {1,2,\ldots,m}$), where $i \in \{1, 2, \ldots, m\}$. As a result, each training sample activates only a subset of neurons, leading to a \textbf{pyramid-shaped reconstruction pattern}: neurons with smaller bias terms tend to be activated by many samples, whereas those with larger bias terms are activated by a few. The adversary then reconstructs individual samples by recursively computing the differences between neighboring neurons, thereby forming $m$ reconstruction bins:
\begin{equation}
    \mathrm{bin}_i=\frac{\nabla_{\mathbf{w}_i}\mathcal{L}-\nabla_{\mathbf{w}_{i-1}}\mathcal{L}}{\nabla_{b_i}\mathcal{L}-\nabla_{b_{i-1}}\mathcal{L}}
\end{equation}

Rigorous analysis demonstrates that when the number of local training samples $d$ is smaller than the number of bins $m$, each sample can be analytically reconstructed within a distinct bin. Accordingly, we have the following attack insight:
\begin{tcolorbox}[colback=gray!10, colframe=gray!50]
\refstepcounter{insight}\label{insight: linear-leakage}
\textbf{Insight 2}: With gradient access, an adversary can manipulate linear-layer neurons to form \textit{multiple reconstruction bins}, enabling batched input recovery with the help of an auxiliary dataset.
\end{tcolorbox}

However, for federated language model finetuning, the \textit{absence of raw gradients} prevents the attacks from the first step. In practice, the local model update is the cumulative training efforts of \textit{many training steps}. When the SGD optimizer is used, this issue can be partially mitigated, as the single-step gradients can be approximately estimated by dividing the final model update by the number of training steps. However, for the more commonly used Adam/AdamW optimizers, the single-step gradients are not accessible, and the same final model update can even result from different training trajectories. This is because Adam and AdamW are stateful optimizers that rely on historical moment estimates to compute subsequent updates \cite{kingma2014adam, loshchilov2019decoupled}. Consequently, reversing their optimization process requires access to detailed intermediate states, which are unavailable under federated finetuning. 

\begin{figure}[t]
    \centering
    \includegraphics[width=0.47\textwidth]{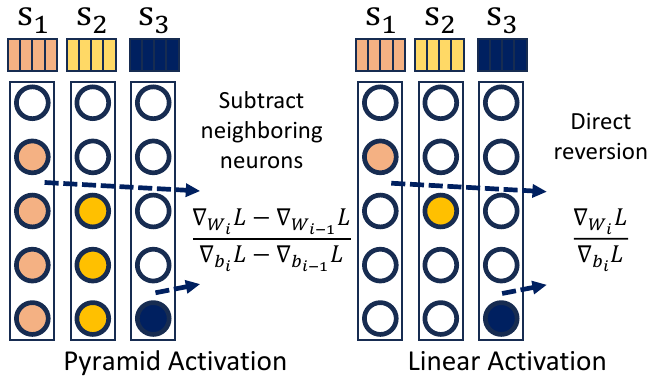}
    \caption{\textit{\sysname} activation pattern: Transforming from pyramid activation to linear activation to adapt for the Adam/AdamW optimizer. }
    \label{fig: activation-pattern} 
\end{figure}

The only exception is the \textit{first training step}, where no previous optimization state exists, and the model update can be approximately reversed as the learning rate multiplied by the sign of the gradient, i.e., \( -\eta\, \mathrm{sign}(\nabla\mathcal{L}) \). This observation motivates us to design reconstruction bins that are activated by only a \textit{single sample} throughout the entire training process, referred to as \textbf{linear activation}, as illustrated in Fig. \ref{fig: activation-pattern}. Unlike the conventional pyramid-shaped activation pattern, which co-activates multiple samples in shared neurons, linear activation avoids mixing and keeps the final update invertible under Adam/AdamW. In the following sections, we first introduce our detailed design to achieve this, followed by rigorous mathematical analysis that explains how and why it works.




\begin{figure*}[th]
    \centering
    \includegraphics[width=1.0\textwidth]{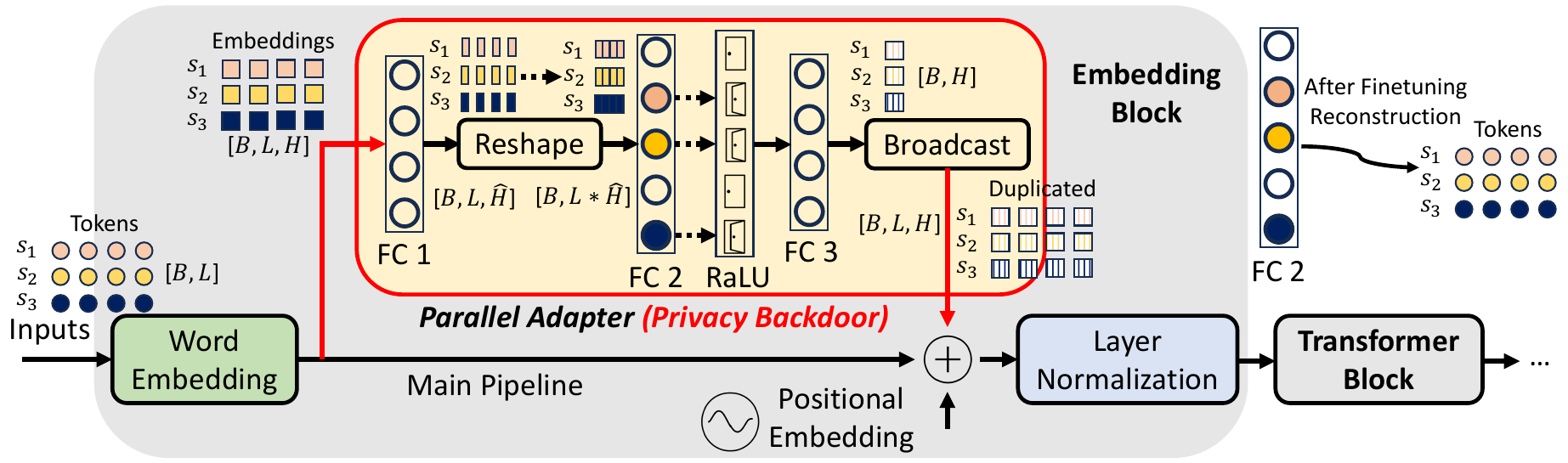}
    \caption{\textit{\sysname} detailed attack flow. Our privacy backdoor is crafted from a legitimate parallel adapter. \textit{\sysname} only modifies the embedding block, leaving the transformer blocks and their adapters unchanged. }
    \label{fig: attack-flow} 
\end{figure*}

\subsection{Backdoor Construction Details}

We illustrate the detailed architecture of the privacy backdoor $\Delta_{\mathrm{adv}}$ in Fig. \ref{fig: attack-flow}. In our design, $\Delta_{\mathrm{adv}}$ is crafted from a benign parallel adapter. The input $\mathbf{x} \in \mathbb{R}^{b\times l \times h}$ to $\Delta_{\mathrm{adv}}$ are the batched embedding vectors of the training samples, which contain the complete training information. The backdoor outputs $\mathbf{o} \in \mathbb{R}^{b\times l \times h}$ with the same dimension as input. But its values are deliberately kept uniform and small across token-index and hidden dimensions, ensuring that $\Delta_{\mathrm{adv}}$ remains functionally invisible and does not interfere with the main training process. Within $\Delta_{\mathrm{adv}}$, there are four key components, and we introduce them one-by-one as follows:

\textbf{Projection Layer:} The first linear layer $L_1$ projects the input embeddings into a lower-dimensional space, $\hat{\mathbf{x}} = L_1(\mathbf{x})$, where $\hat{\mathbf{x}} \in \mathbb{R}^{b\times l \times \hat{h}}$ and $\hat{h} \ll h$, effectively reducing dimensionality and minimizing computational overhead. This is crucial because, after reshaping, the input size of the subsequent linear layer $L_2$ becomes $l \times \hat{h}$ and would grow excessively if the original dimension $h$ were used. Technically, we use principal component analysis (PCA) to retain only the principal components that preserve the majority of the variance in the original embeddings:
\begin{equation}
    \hat{\mathbf{x}}=(\mathbf{x}-\mu)\mathbf{W}_{\mathrm{T}}
\end{equation}
where $\mu$ is the mean vector and $\mathbf{W}_{\mathrm{T}}\in \mathbb{R}^{h\times \hat{h}}$ refers to the projection matrix that preserves top-$\hat{h}$ eigenvectors generated by the model's original word embedding matrix $\mathbf{W}_{\mathrm{E}}\in\mathbb{R}^{v\times h}$. As a reference, in the BERT model, the original hidden dimension $h=768$ and reduced dimension $\hat{h}=64$. Such a setting ensures that the dimension is significantly reduced, while also preserving enough information for accurate embedding-to-token reversion.

\textbf{Memorization Layer:} The second linear layer $L_2$ plays a key role in “memorizing” gradients, which can be reverted to the training samples according to Ins. \ref{insight: gradient-invert} and Ins. \ref{insight: linear-leakage}. Before entering $L_2$, we reshape $\hat{\mathbf{x}}\in \mathbb{R}^{b\times l \times h}$ into $\hat{\mathbf{x}}\in \mathbb{R}^{b\times l \cdot h}$ to fit the dimension of $L_2$. This reshaping ensures that all token gradients are collectively “memorized” within \textit{a single neuron}, which helps to separate batched samples during reconstruction (one neuron, one sample).

To construct the weight $\mathbf{W}_2$ and bias $\mathbf{b}_2$ of layer $L_2$, we adopt a similar method described in Section 4.1. We keep the row vector $\mathbf{w}_2^i$ identical to a random vector $\mathbf{r}_2$ across all neurons in $L_2$ and feed the auxiliary dataset $D_{\mathrm{aux}}$ through the model up to this layer to derive the overall bias distribution $\psi$ of $\mathbf{b}_2$. 
$\psi$ covers all possible values of $\mathbf{b}_2$ and we further divide this distribution into $m$ intervals with the same probability, i.e., $B_k=[F^{-1}(\frac{k}{m}),\ F^{-1}(\frac{k+1}{m}))$, where $F$ refers to the cumulative density function (CDF) of $\psi$, $k\in \{0, 1,\cdots,m-1\}$, and $m$ indicates the adversary's maximum reconstruction capability. Such a partition ensures that at most one sample from the dataset falls into each interval. Accordingly, the adversary can construct $\mathbf{b}_2$ to form $m$ continuous intervals arranged in ascending order as follows:
\begin{equation}
\label{equ: layer2-design}
\begin{aligned}
    \mathbf{W}_2&=[\mathbf{r}_2, \mathbf{r}_2, \cdots, \mathbf{r}_2]^T \\
    \mathbf{b}_2&=-[F^{-1}(\frac{1}{m}), F^{-1}(\frac{2}{m}), \cdots, F^{-1}(1)]^T
\end{aligned}
\end{equation}

\textbf{Ranged Linear Unit:} Eq. \ref{equ: layer2-design} has constructed the second layer $L_2$ to form $m$ intervals. To further convert them into reconstruction bins, a non-linear activation function is needed to truncate and trap the gradients. The existing ReLU activation function progressively activates neurons from the lowest upward when new training samples arrive, forming a pyramid-shaped activation pattern as illustrated in Fig. \ref{fig: activation-pattern}. However, as we have discussed, this pyramid-shaped activation within each training batch causes continuous updates to the lower neurons, altering both the corresponding weight vectors $\mathbf{w}_2^i$ and bias terms $b_2^i$ that define the interval boundaries. Moreover, the accumulated training effects introduced by Adam/AdamW make per-step gradients effectively unrecoverable, causing cross-step reconstruction interference. 

To address this, we propose a \textit{new activation function} named Ranged Linear Unit, or \textit{RaLU}. Compared to ReLU, which activates all inputs that are larger than zero, RaLU sets a configurable upper bound to further restrict the activation area. For each neuron in the second layer, we set the upper bound as the corresponding interval length:
\begin{equation}
    \mathrm{RaLU}(z_i)=
    \begin{cases}
    z_i, & 0<z_i<F^{-1}(\frac{i+1}{n})-F^{-1}(\frac{i}{n}), \\
    0, & \text{otherwise.}
    \end{cases}
\end{equation}
where $z_i$ refers to the input to the $i^{th}$ neuron and $i\in\{0,1,\cdots,m-1\}$.

This design ensures that each sample activates a unique neuron, forming a linear activation pattern. Under this pattern, each neuron is updated only \textit{once by a single sample} during the training run, preventing gradient or parameter accumulation. Consequently, the gradient information can be directly recovered from the final parameter updates of $L_2$, even under Adam or AdamW optimizers, since a single-step update is straightforward to invert and does not depend on intermediate optimizer states.

In summary, we define the reconstruction bins as the combination of the second linear layer and the RaLU activation function. Each bin will “memorize” at most one gradient, which can later be reverted back to training samples using Eq. \ref{equ: gradient-invert}. We also identify the following attack insight:

\begin{tcolorbox}[colback=gray!10, colframe=gray!50]
\refstepcounter{insight}\label{insight: ralu-activation}
\textbf{Insight 3}: Activation functions that constrain activation ranges, such as RaLU, can force \textit{each sample} to \textit{activate a single, unique neuron} during finetuning, effectively mitigating cross-sample and cross-step reconstruction interference. 
\end{tcolorbox}


\textbf{Output Layer:} The final component of the privacy backdoor is the output linear layer $L_3$, which maps the intermediate representations back to the original embedding dimension $h$ and broadcasts them across token indexs to produce the output $\mathbf{o} \in \mathbb{R}^{b\times l\times h}$, matching the input shape.

For its weights $\mathbf{W}_3$ and bias $\mathbf{b}_3$, we initialize the row vectors $\mathbf{w}_3^j$ ($j\in\{1,2,\cdots,h\}$) identical to a random vector $\mathbf{r}_3$ and all entries of bias $\mathbf{b}_3$ constant, i.e., $\mathbf{W}_3=[\mathbf{r}_3, \mathbf{r}_3, \cdots, \mathbf{r}_3]^T$ and $\mathbf{b}_3=[0, 0, \cdots, 0]^T$. Based on this, the output $\mathbf{o} \in \mathbb{R}^{b\times l\times h}$ has \textit{identical values} across the token-index ($l$) and hidden ($h$) dimensions, but differs across the batch ($b$) dimension. This property causes the contribution of output $\mathbf{o}$ to be canceled out during layer normalization, effectively nullifying the influence of the privacy backdoor module during training/finetuning. We present this attack insight as follows, and provide detailed mathematical explanations in Section 4.4.

\begin{tcolorbox}[colback=gray!10, colframe=gray!50]
\refstepcounter{insight}\label{insight: layernorm}
\textbf{Insight 4}: 
Backdoor outputs that remain constant across token-index and hidden dimensions can be \textit{fully canceled out} by LayerNorm. Ensuring that the backdoor is invisible during model fine-tuning.
\end{tcolorbox}

In summary, we illustrate the mathematical details of the privacy backdoor $\Delta_{\mathrm{adv}}$ in Tab. \ref{tab: backdoor-construction}. 

\begin{table}[t]
\caption{\textit{\sysname} detailed backdoor construction steps.}
\label{tab: backdoor-construction}
\normalsize
\centering
\renewcommand{\arraystretch}{1.08} 
\begin{tabular}{|p{0.95\linewidth}|}
\hline
\textbf{\textsf{\sysname} Privacy Backdoor Construction} \\
\hline

\textbf{Backdoor Placement}: 
The privacy backdoor $\Delta_{\mathrm{adv}}$ is placed after the word embedding module and before the LayerNorm module, same as a parallel adapter. \\[2pt]

\textbf{Hyper Parameters}: 
Hidden dim $h$, reduced hidden dim $\hat{h}$, reconstruction bin size $m$, and max sequence length $l$. \\[2pt]

\textbf{Backdoor Components:} \\[2pt]

\quad \textbf{Projection Layer $L_1$:} 
$\mathbf{x}\in\mathbb{R}^{b\times l \times h}\rightarrow\hat{\mathbf{x}}\in\mathbb{R}^{b\times l \times \hat{h}}$ \\[1pt]
\quad\quad Take the embed matrix $\mathbf{E}=[\mathbf{e}_1, \mathbf{e}_2, \cdots, \mathbf{e}_v]^T$ \\
\quad\quad Compute $\boldsymbol{\mu}=\frac{1}{v}\sum_{i=1}^v \mathbf{e}_i$, 
$\mathbf{E}-\mathbf{1}\boldsymbol{\mu}^T=\mathbf{U}\Sigma\mathbf{V}^T$ \\
\quad\quad Select top $\hat{h}$ components $\mathbf{W}_\mathrm{T}=\mathbf{V}_{\hat{h}}$ \\
\quad\quad Formalize $\hat{\mathbf{x}}=L_3(\mathbf{x})=(\mathbf{x}-\mu)\mathbf{W}_{\mathrm{T}}$ \\[2pt]

\quad \textbf{Data Reshape:} 
$\hat{\mathbf{x}}\in\mathbb{R}^{b\times l \times h}\rightarrow\hat{\mathbf{x}}\in\mathbb{R}^{b\times l \cdot \hat{h}}$ \\[2pt]

\quad \textbf{Memorization Layer $L_2$:} 
$\hat{\mathbf{x}}\in\mathbb{R}^{b\times l \cdot \hat{h}}\rightarrow \mathbf{z}\in\mathbb{R}^{b\times m}$ \\[1pt]
\quad\quad Assign $\mathbf{W}_2=[\mathbf{r_2}, \mathbf{r_2}, \cdots, \mathbf{r_2}]^T$, $\mathbf{r_2}=\mathrm{rand}(l \cdot\hat{h})$ \\
\quad\quad Estimate $b_2^i\sim \psi$ with $D_{\mathrm{aux}}$ and $F(t)=\mathrm{P}\{b_2^i\le t\}$ \\
\quad\quad Assign $\mathbf{b}_2=-[F^{-1}(\frac{1}{m}), F^{-1}(\frac{2}{m}), \cdots, F^{-1}(1)]^T$ \\[2pt]

\quad \textbf{RaLU Activation:} 
$\mathbf{z}\in\mathbb{R}^{b\times m}\rightarrow\mathbf{z}\in\mathbb{R}^{b\times m}$ \\
\quad\quad For $z_i$, $i\in \{1,2,\cdots,m\}$: \\
\quad\quad\quad If $0<z_i<F^{-1}(\frac{i+1}{n})-F^{-1}(\frac{i}{n})$: $\mathrm{RaLU}(z_i)=z_i$ \\
\quad\quad\quad Else: $\mathrm{RaLU}(z_i)=0$ \\[2pt]

\quad \textbf{Output Layer:} 
$\mathbf{z}\in\mathbb{R}^{b\times m}\rightarrow \mathbf{o}\in \mathbb{R}^{b\times h}$ \\[1pt]
\quad\quad Assign  $\mathbf{W}_3=[\mathbf{r_3}, \mathbf{r_3}, \cdots, \mathbf{r_3}]^T$, $\mathbf{r_3}=\mathrm{rand}(m)$ \\
\quad\quad Assign $\mathbf{b}_3=\mathbf{0}$ \\[2pt]

\quad \textbf{Data Broadcast:} 
$\mathbf{o}\in\mathbb{R}^{b\times h}\rightarrow \mathbf{o}\in \mathbb{R}^{b\times t\times h}$ \\
\hline
\end{tabular}
\end{table}

\subsection{Closed-form Reversion}

We consider that after model finetuning, the memorization layer $L_2$'s parameters are $\tilde{\mathbf{W}}_2$ and $\tilde{\mathbf{b}}_2$. The embedding vectors of training samples $\tilde{\mathbf{x}}\in \mathbb{R}^{m\times t\cdot \hat{h}}$ can be reversed as follows:
\begin{equation}
    \label{equ: closed-form reversion}
    \begin{aligned}
        \tilde{\mathbf{x}}=\frac{\tilde{\mathbf{W}}_2-\mathbf{W}_2}{\tilde{\mathbf{b}}_2-\mathbf{b}_2}
    \end{aligned}
\end{equation}

When the SGD optimizer is used, because all neurons of $L_2$ is designed to update \textit{at most once in one epoch}, all non-zero activated/updated neurons (index by $j$) of $L_2$ satisfy $\tilde{\mathbf{W}}_2^j-\mathbf{W}_2^j=-\eta\nabla_{\mathbf{W}_2^j}\mathcal{L}$ and $\tilde{\mathbf{b}}_2^j-\mathbf{b}_2^j=-\eta\nabla_{\mathbf{b}_2^j}\mathcal{L}$. This makes Eq. \ref{equ: closed-form reversion} identical to Eq. \ref{equ: gradient-invert}, indicating that the corresponding sample $\tilde{x}_j$ can be \textit{perfectly reconstructed}. For the more complex Adam/AdamW optimizer, the single-step model update can be reverted as $\tilde{\mathbf{W}}_2^j-\mathbf{W}_2^j=-\eta\mathrm{sign}(\nabla_{\mathbf{W}_2^j}\mathcal{L})$ and $\tilde{\mathbf{b}}_2^j-\mathbf{b}_2^j=-\eta\mathrm{sign}(\nabla_{\mathbf{b}_2^j}\mathcal{L})$, where $\mathrm{sign}$ refers to the direction of vectors. This indicates that there is a reversion information loss (although slight) when the Adam/AdamW optimizer is used, and Eq. \ref{equ: closed-form reversion} can only achieve \textit{approximated reconstruction}. 
More detailed mathematical explanations for how this inversion is derived can be found in Appendix B. Based on this, we discover the following attack insight:

\begin{tcolorbox}[colback=gray!10, colframe=gray!50]
\refstepcounter{insight}\label{insight: optimizers}
\textbf{Insight 5}: 
SGD exposes raw gradients, enabling \textit{exact sample inversion}. In comparison, Adam/AdamW obscures them with momentum and adaptive states, leaving smoothed gradients and only allowing \textit{approximate reconstruction}.
\end{tcolorbox}

\subsection{Bypassing Secure Aggregation}

Based on our previous design, \textit{\sysname} can reverse a union set of local finetuning datasets $\cup_{i=1}^n D_i$. However, the secure aggregation can still ensure “privacy-by-shuffling”, and prevent the adversary from attributing the reconstructed samples to individual clients. To further enable targeted attack, we keep the backdoor unchanged for the victim, denoted as $\Delta_{\mathrm{adv}}^{\mathrm{target}}$, and slightly change the privacy backdoor sent to other clients, denoted by $\hat{\Delta}_{\mathrm{adv}}^{\mathrm{others}}$. The difference between $\hat{\Delta}_{\mathrm{adv}}^{\mathrm{others}}$ and $\Delta_{\mathrm{adv}}^{\mathrm{target}}$ is that we change the bias vector of the second layer $\mathbf{b}_2$ to be $\mathbf{\hat{b}}_2=-[F^{-1}(1-\epsilon), F^{-1}(1-\epsilon), \cdots, F^{-1}(1-\epsilon)]^T$, where $\epsilon$ refers to a small constant value. Because $F^{-1}(1-\epsilon)$ has already reached the boundary of the whole distribution $\psi$, this design ensures that almost \textit{no neuron is activated} during the local fine-tuning process. This makes the model updates negligible for all other clients except the victim, exposing the individual model update of the client, which can be inverted back to training samples via Eq. \ref{equ: closed-form reversion}. Mathematically, this can be expressed as:
\begin{equation}
    \theta^{r+1}=\frac{1}{n}\sum_{i=1}^n u_i= \frac{1}{n}\sum_{i=1}^n \theta_i=\theta^r+\frac{1}{n}(\theta_{\mathrm{victim}}-\theta^r)
\end{equation}

\subsection{Analysis}

We highlight the following properties of our proposed attack.

\textbf{Zero Performance Impact}: Suppose the original word and positional embeddings are $\mathbf{e}_{w}$, and $\mathbf{e}_{p}$. Without any backdoor, the $LayerNorm$ can be expressed as:

\begin{equation}
    \mathrm{LayerNorm}(\mathbf{z}) 
= \gamma \odot \frac{\mathbf{z} - \boldsymbol{\mu}}{\sqrt{\boldsymbol{\sigma}^2 + \epsilon}} + \beta
\end{equation}
where $\mathbf{z} = \mathbf{e}_w + \mathbf{e}_p$ denotes the original input embeddings, $\boldsymbol{\mu}=\frac{1}{h}\sum_{i=1}^{h}\mathbf{z}_i$, $\boldsymbol{\sigma}^2=\frac{1}{h}\sum_{i=1}^h (\mathbf{z}_i-\boldsymbol{\mu})^2$ refers to feature-level mean and variance, and $\gamma$, $\beta$ refers to learnable parameters. With the backdoor inserted, the input to \text{LayerNorm} becomes $\hat{\mathbf{z}} = \mathbf{z} + \mathbf{o}=[\mathbf{z_1}+\mathbf{v}, \mathbf{z_2}+\mathbf{v}, \cdots, \mathbf{z_h}+\mathbf{v}]^T$, where  $\mathbf{v}$ refers to the constant feature vector of $\mathbf{o}$. Accordingly, we can calculate $\boldsymbol{\mu}_{\hat{\mathbf{z}}}=\frac{1}{h}\sum_{i=1}^{h}\mathbf{z}_i+\mathbf{v}$, and $\boldsymbol{\sigma}_{\hat{\mathbf{z}}}^2=\frac{1}{h}\sum_{i=1}^h (\hat{\mathbf{z}}-\boldsymbol{\mu}_{\hat{\mathbf{z}}})^2=\frac{1}{h}\sum_{i=1}^h(\mathbf{z}_i+\mathbf{v}-\boldsymbol{\mu}-\mathbf{v})^2=\frac{1}{h}\sum_{i=1}^h (\mathbf{z}_i-\boldsymbol{\mu})^2=\boldsymbol{\sigma}^2$. Take a step further, we can derive:
\begin{equation}
    \begin{aligned}
    \mathrm{LayerNorm}(\hat{\mathbf{z}})&=\gamma \odot \frac{\hat{\mathbf{z}} - \boldsymbol{\mu}_{\hat{\mathbf{z}}}}{\sqrt{\boldsymbol{\sigma_{\hat{\mathbf{z}}}}^2 + \epsilon}} + \beta\\
    &=\gamma \odot \frac{\mathbf{z}+\mathbf{o} - \boldsymbol{\mu}_{\hat{\mathbf{z}}}}{\sqrt{\boldsymbol{\sigma}^2 + \epsilon}} + \beta\\
    &=\mathrm{LayerNorm}(\mathbf{z})
    \end{aligned}
\end{equation}

This implies that the LayerNorm produces identical outputs with or without our privacy backdoor. Consequently, the subsequent transformer blocks, which are the core components of language models, perceive no difference, indicating that our attack module incurs negligible impact on training performance.

\textbf{Scalable Reconstruction:}
The adversary's reconstruction capability is largely defined by the number of reconstruction bins $m$. In theory, one reconstruction bin can reconstruct one finetuning sample, and the adversary can select $m$ in the number of a few thousand (e.g., 4096), as this is the typical size of linear layers in language models. Under this setting, our attack can reconstruct thousands of samples. For each sample, we set a maximum token length $l$, adjustable by the attacker according to the targeted content. In practice, the adversary can select $l$ up to a few hundred (e.g., 384) to reverse long sequences. Note that both the number of bins $m$ and the maximum token length $l$ are \textit{\textbf{attacker-controlled parameters}}, and can be scaled as needed. However, a larger memorization layer incurs higher memory overhead and cannot be scaled indefinitely. We provide further discussion for this scalability--memory space trade-off in Section \ref{sec: discussion}.


\textbf{Adaptive to LoRA Finetuning:} 
LoRA fine-tuning has been widely adopted for adapting transformer blocks with low overhead. Compared to full-rank MLP-style \emph{parallel} or \emph{serial} adapters, LoRA parameterizes the update to a target weight matrix $\mathbf{W}_{\mathrm{origin}}\in \mathbb{R}^{m\times s}$ using a low-rank decomposition, i.e., $\mathbf{W}_{\mathrm{new}}=\mathbf{W}_{\mathrm{origin}}+\Delta\mathbf{W}$ with $\Delta\mathbf{W}=\mathbf{B}\mathbf{A}$, where $\mathbf{B}\in\mathbb{R}^{m\times q}$, $\mathbf{A}\in \mathbb{R}^{q\times s}$, and $q \ll m,s$. Consequently, the update $\Delta\mathbf{W}$ has rank at most $q$, leading to significantly lower computation and storage overhead than full-rank adapter variants.


For \textit{\sysname}, LoRA can be applied to all transformer blocks, and the privacy backdoor remains effective as long as the gradients are back-propagated through it. However, LoRA is not suitable for the backdoor itself, particularly for the critical memorization layer. This is because the inevitable information loss from low-rank approximation degrades the fidelity of the stored gradients, thereby reducing the attack’s effectiveness.




\section{Evaluation}
\label{sec: evaluation}

\begin{figure*}[t]
  \centering
  \begin{tcolorbox}[colback=gray!15, colframe=gray!40, boxrule=0.4pt, 
                    width=\linewidth, left=6pt, right=6pt, top=4pt, bottom=4pt,
                    sharp corners, enhanced]
    \textbf{Original Text}:\\[2pt]
    \textbf{Context:} Beyoncé announced a hiatus from her music career in January 2010, heeding her mother's advice, ``to live life, to be inspired by things again''. During the break she and her father parted ways as business partners. Beyoncé's musical break lasted nine months and saw her visit multiple European cities, the Great Wall of China, the Egyptian pyramids, Australia, English music festivals, and various museums and ballet performances.\\
    \textbf{Question:} Which famous landmark did Beyoncé see in China?\\
    \textbf{Answer:} the Great Wall of China 

    \vspace{6pt}
    \textbf{Reconstructed Text 1 (SGD)}:\\[2pt]
    \textbf{Context:} \textcolor{blue}{Beyoncé announced a hiatus from her music career in January 2010, heeding her mother's advice, ``to live life, to be inspired by things again''. During the break she and her father parted ways as business partners. Beyoncé's musical break lasted nine months and saw her visit multiple European cities, the Great Wall of China, the Egyptian pyramids, Australia, English music festivals, and various museums and ballet performances. }\\
    \textbf{Question:} \textcolor{blue}{Which famous landmark did Beyoncé see in China?} \\
    \textbf{Answer:} \textcolor{blue}{the Great Wall of China}

    \vspace{6pt}
    \textbf{Reconstructed Text 2 (AdamW)}:\\[2pt]
    \textbf{Context:} Beyoncy \textcolor{red}{announced} several \textcolor{red}{hiatus} beside hers \textcolor{red}{music career} onto \textcolor{red}{January 2010}, heed hers \textcolor{red}{mother’s advice}, \textcolor{red}{``to live life, to become inspired by things again''.} \textcolor{red}{During} hers \textcolor{red}{break} herself argues \textcolor{red}{hers father} parted routes since \textcolor{red}{business partners}. \textcolor{red}{Beyoncy’s musical break lasted nine months} argues saw hers \textcolor{red}{visit multiple European cities}, \textcolor{red}{the Great Wall} towards \textcolor{red}{China}, the \textcolor{red}{Egyptian pyramids}, \textcolor{red}{Australia}, \textcolor{red}{English music festivals} argues various \textcolor{red}{museums} argues \textcolor{red}{ballet performances}. \\
    \textbf{Question:} \textcolor{red}{Which famous landmark did} Beyoncy look beside \textcolor{red}{China?}\\
    \textbf{Answer:} \textcolor{red}{the Great Wall of China}

  \vspace{6pt}
    \textbf{LLM Rephrased (AdamW)}:\\[2pt]
    \textbf{Context:} \textcolor{brown}{Beyoncé announced a hiatus from her music career in January 2010, following her mother’s advice to “live life and become inspired by things again.” During her break, she and her father ended their professional relationship as business partners. Her musical hiatus lasted for nine months, during which she traveled extensively—visiting several European cities, the Great Wall of China, the Egyptian pyramids, Australia, various English music festivals, and numerous museums and ballet performances. }\\
    \textbf{Question:} \textcolor{brown}{Which famous landmark did Beyoncé visit in China?} \\
    \textbf{Answer:} \textcolor{brown}{the Great Wall of China}
  \end{tcolorbox}
  \caption{\textit{\sysname} reconstruction examples on the SQuAD dataset. 
  Text 1 corresponds to fine-tuning with SGD; Text 2 corresponds to fine-tuning with AdamW. SGD yields nearly perfect reconstruction, while AdamW leads to approximated recovery with minor distortions. To improve their readability and semantic coherence, the adversary could further refine the reconstructed samples by leveraging LLMs to rephrase the content.
  More examples are provided in Appendix D.}
  \label{fig: reconstruction-examples}
\end{figure*}

\begin{table*}[t]
    \centering
    \caption{The general reconstruction performance of \textit{\sysname} on different models and datasets.}
    \vspace{2pt}
    \small
    \setlength{\tabcolsep}{4pt}
    \begin{tabular}{lcccccccc}
        \toprule
        \multirow{2}{*}{\textbf{Dataset}} &
        \multirow{2}{*}{\textbf{Token Len.}} &
        \multirow{2}{*}{\textbf{Optimizer}} &
        \multicolumn{1}{c}{\textbf{BERT}} &
        \multicolumn{1}{c}{\textbf{GPT-2}} &
        \multicolumn{1}{c}{\textbf{Qwen2-1.5B}} &
        \multicolumn{1}{c}{\textbf{Llama3-3B}} \\
        \cmidrule(r){4-7}
        & & & Rate$|$Sim.$\pm$Std & Rate$|$Sim.$\pm$Std & Rate$|$Sim.$\pm$Std & Rate$|$Sim.$\pm$Std \\
        \midrule
        \multirow{2}{*}{AGNews}
            & \multirow{2}{*}{128}
            & SGD   & 0.774$|$0.994$\pm$0.028 & 0.665$|$0.990$\pm$0.050 & 0.714$|$0.997$\pm$0.025 & 0.700$|$0.992$\pm$0.057 \\
            &       & AdamW & 0.746$|$0.767$\pm$0.090 & 0.744$|$0.779$\pm$0.115 & 0.733$|$0.830$\pm$0.102 & 0.627$|$0.817$\pm$0.085 \\
        \midrule
        \multirow{2}{*}{SQuAD}
            & \multirow{2}{*}{384}
            & SGD   & 0.748$|$0.979$\pm$0.090 & 0.642$|$0.988$\pm$0.071 & 0.674$|$0.991$\pm$0.061 & 0.750$|$0.997$\pm$0.032 \\
            &       & AdamW & 0.772$|$0.714$\pm$0.097 & 0.734$|$0.682$\pm$0.119 & 0.531$|$0.668$\pm$0.112 & 0.724$|$0.926$\pm$0.072 \\
        \midrule
        \multirow{2}{*}{mSQuAD}
            & \multirow{2}{*}{384}
            & SGD   & 0.796$|$0.955$\pm$0.119 & 0.688$|$0.963$\pm$0.106 & 0.776$|$0.959$\pm$0.121 & 0.750$|$0.965$\pm$0.106 \\
            &       & AdamW & 0.751$|$0.801$\pm$0.078 & 0.739$|$0.806$\pm$0.078 & 0.736$|$0.767$\pm$0.077 & 0.723$|$0.845$\pm$0.059 \\
        \midrule
        \multirow{2}{*}{GSM8K}
            & \multirow{2}{*}{512}
            & SGD   & -- & 0.608$|$0.978$\pm$0.100 & 0.706$|$0.959$\pm$0.142 & 0.610$|$0.946$\pm$0.157 \\
            &       & AdamW & -- & 0.654$|$0.524$\pm$0.116 & 0.595$|$0.663$\pm$0.084 & 0.639$|$0.865$\pm$0.075 \\
        \bottomrule
    \end{tabular}
    \\
    \vspace{2mm}
    \footnotesize{BERT is excluded from GSM8K because it is an encoder-only model for classification/span extraction.}
    \label{tab: general-performance}
\end{table*}

\begin{figure*}[t]
    \centering
    \subfigure[AGNews/SGD]{\includegraphics[width=0.242\textwidth]{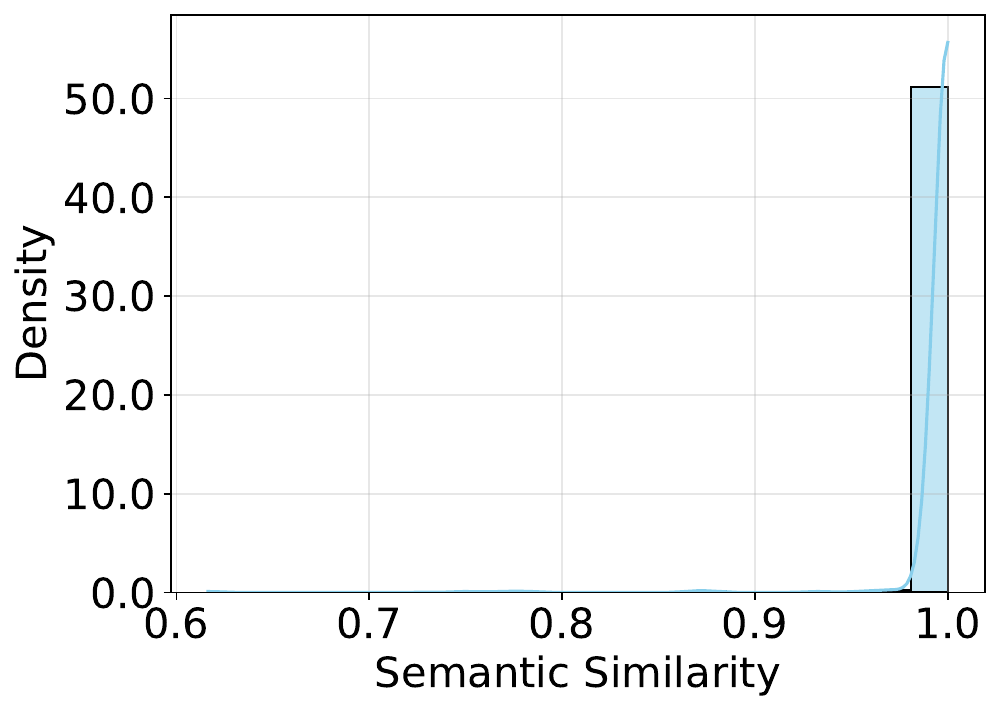}}
    \subfigure[AGNews/AdamW]{\includegraphics[width=0.242\textwidth]{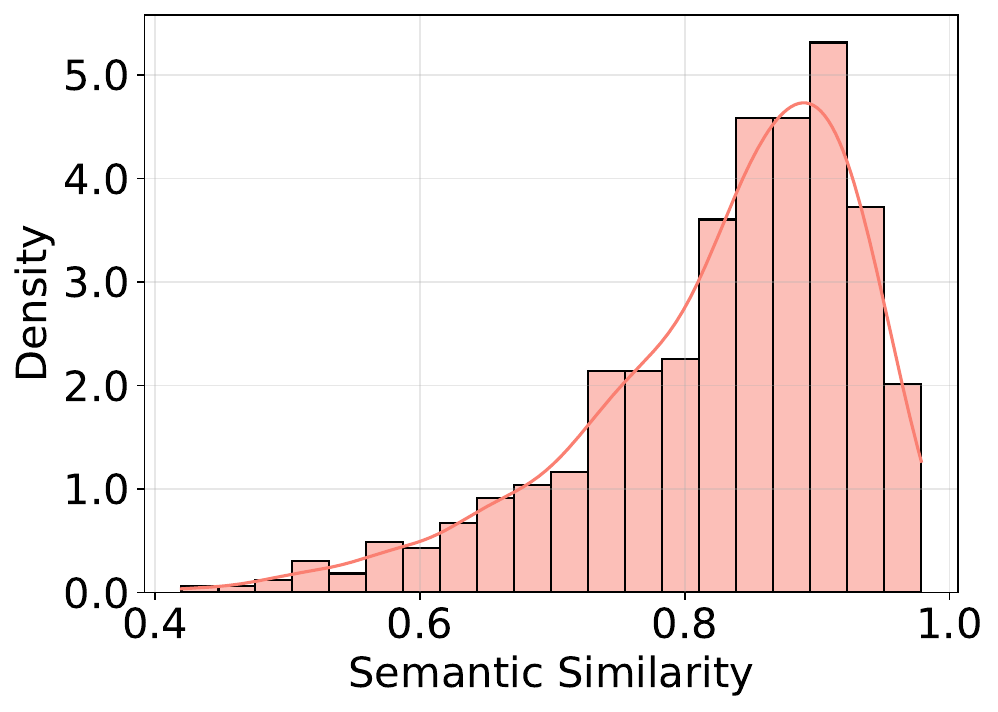}}
    \subfigure[SquAD/SGD]{\includegraphics[width=0.242\textwidth]{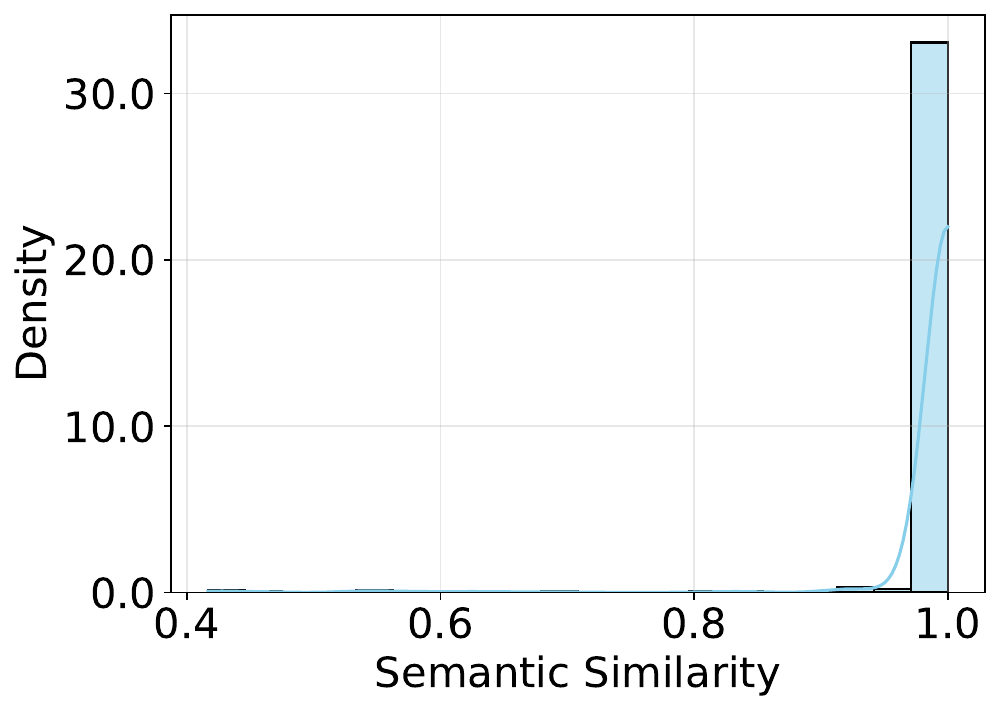}}
    \subfigure[SquAD/AdamW]{\includegraphics[width=0.242\textwidth]{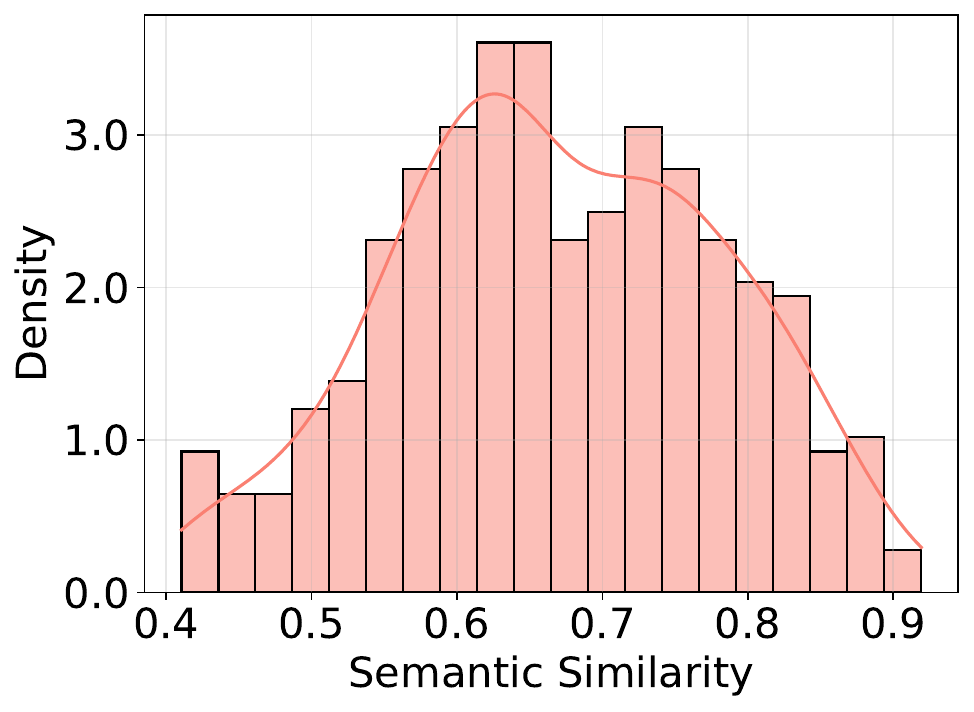}}
    \subfigure[mSQuAD/SGD]{\includegraphics[width=0.242\textwidth]{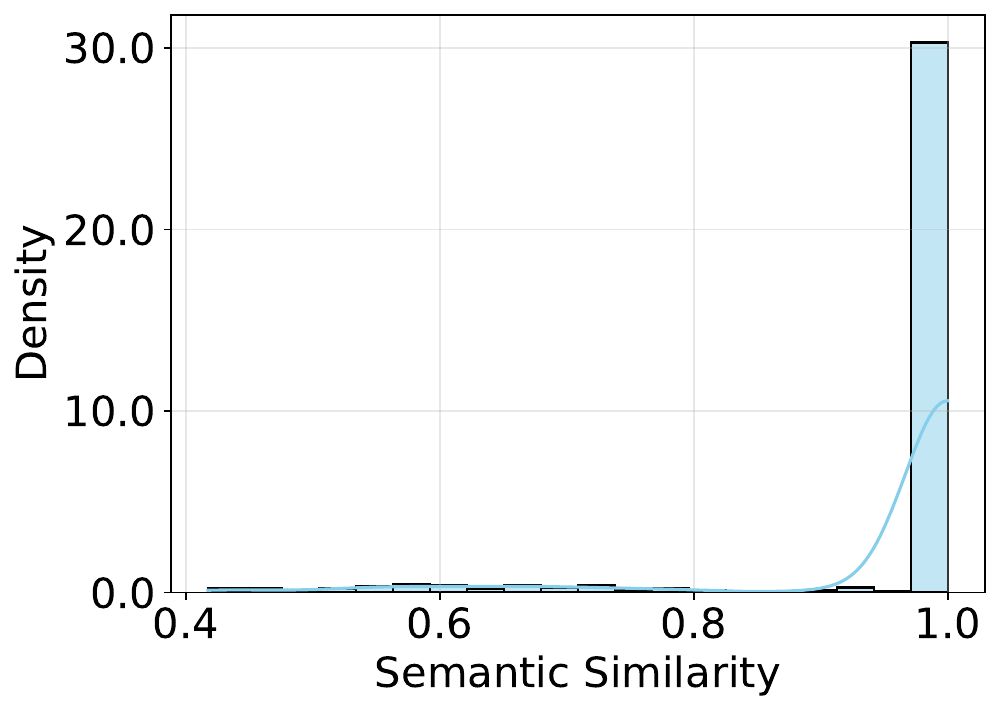}}
    \subfigure[mSQuAD/AdamW]{\includegraphics[width=0.242\textwidth]{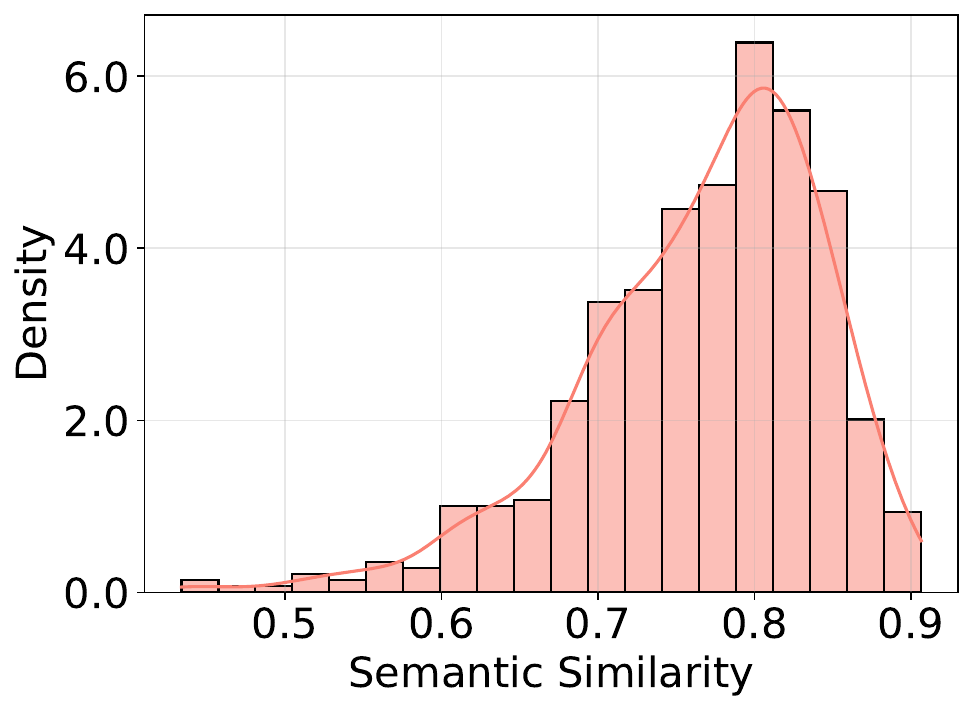}}
    \subfigure[GSM8K/SGD]{\includegraphics[width=0.242\textwidth]{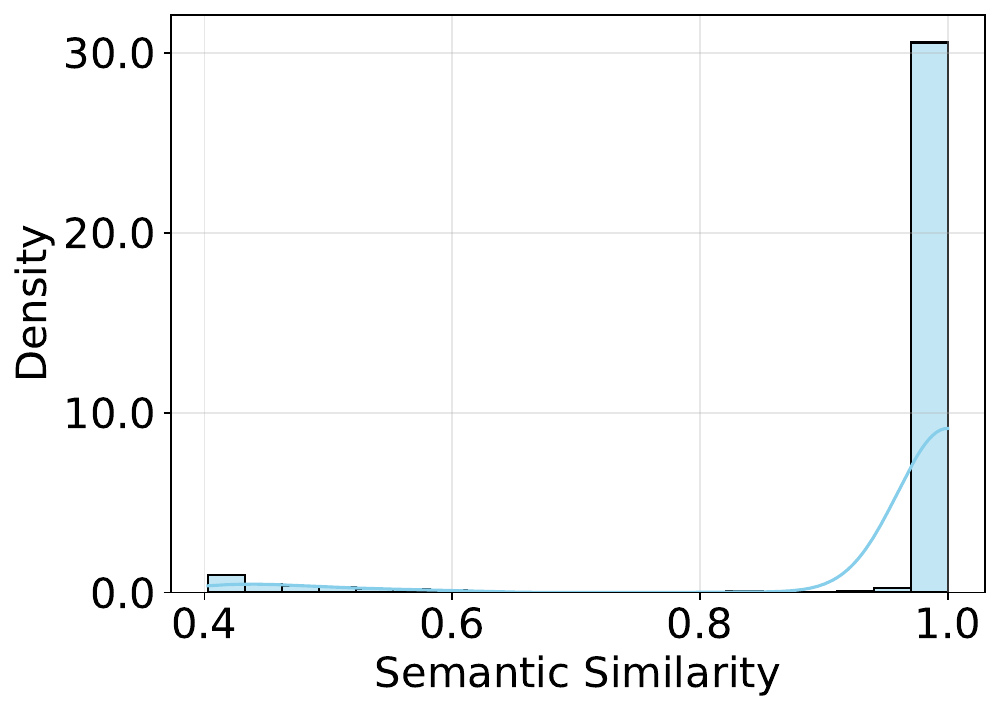}}
    \subfigure[GSM8K/AdamW]{\includegraphics[width=0.242\textwidth]{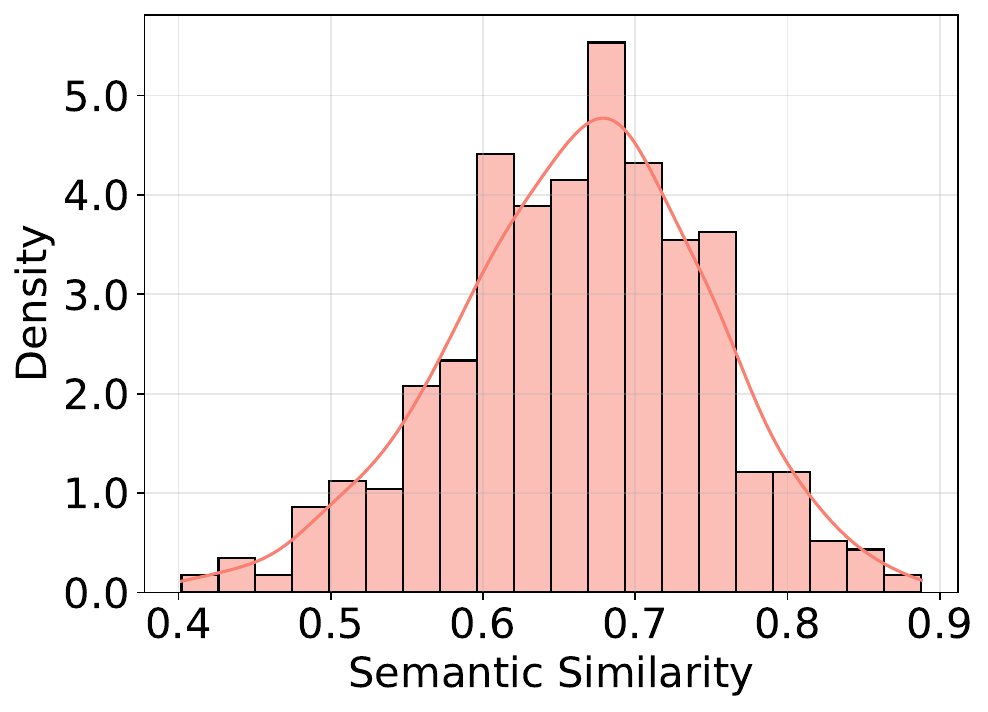}}
    \caption{The semantic similarity distribution of reconstructed samples for Qwen2-1.5B. }
    \label{fig: FL-attack-performance} 
\end{figure*}

\subsection{Experiment Settings}

\textbf{Implementation:} We implemented \textit{\sysname} on the PyTorch platform. We run all the experiments on a server equipped with an AMD EPYC 7643 48-Core CPU Processor, four NVIDIA GeForce A100-SXM4-80GB GPUs, and Ubuntu 22.04.3 LTS. 

\textbf{Models:} We conducted experiments on four language models, including BERT \cite{devlin2019bert}, GPT-2 \cite{radford2019language}, Qwen 2 \cite{bai2023qwen}, and Llama 3.2 \cite{dubey2024llama}. We download their latest pre-trained version from Hugging Face as the base models.

\textbf{Datasets:} We evaluated \textit{\sysname}'s performance on four different datasets, including AGNews \cite{zhang2015character}, SQuAD \cite{rajpurkar2016squad}, EMRQA-mSQuAD \cite{eladio2024emrqa}, and GSM8K \cite{cobbe2021gsm8k}. Their detailed introduction can be found in Appendix C.

\textbf{Evaluation Metrics:} We use two evaluation metrics to evaluate our proposed attack, including semantic similarity (Sim.) and reconstruction rate (Rate). For the semantic similarity, we measure the sentence embedding cosine similarity between the original $\mathbf{x}$ and the reconstructed ones $\tilde{\mathbf{x}}$, which can be represented as:
\begin{equation}
    s_{\mathrm{semantic}}(\mathbf{x}, \tilde{\mathbf{x}})=\cos (\mathrm{Enc}(\mathbf{x}), \mathrm{Enc}(\tilde{\mathbf{x}}))
\end{equation}
where $\mathrm{Enc}$ refers to the semantic encoder, and $\cos()$ refers to the cosine similarity function. In our experiment, we use the widely adopted Sentence-Transformer \cite{reimers2019sentence} as the semantic encoder. The reconstruction rate measures the ratio of reconstruction samples. Note that all the experiment results we report are the vanilla reconstruction performance without any LLM enhancement (as shown in Fig. \ref{fig: reconstruction-examples}). Additional results for LLM refinement can be found in Appendix F.

\textbf{Default FL Settings:} For each dataset, we select 8000 samples from the FL training set, denoted as $D_{\mathrm{train}}$, and 2000 samples from the test set to form the auxiliary dataset, denoted as $D_{\mathrm{aux}}$. This ensures that there is \textit{no overlap} between the two datasets. We assume there are 10 clients in the FL system, and the training dataset $D_{\mathrm{train}}$ is split uniformly across them, with each client holding 800 training samples. We assume the state-of-the-art secure aggregation mechanism is in place \cite{bonawitz2017practical}, and the adversary can only obtain the aggregated model update. We select one client as the attack victim and aim to reconstruct all training samples within its local dataset, denoted by $D_{\mathrm{target}}$. We assume each client conducts local training for 16 training steps with a batch size of 50. The client can use either the SGD or AdamW optimizers to perform the local training. The adversary is assumed to construct $m = 2000$ reconstruction bins for the attack. 

\subsection{Reconstruction Capability Evaluation}

\textbf{General Performance:} We report the experiment results in Tab. \ref{tab: general-performance}. Specifically, we report the reconstruction rate, the average semantic similarity of all reconstructed samples, and the corresponding standard deviation.

From the results, we observe that the reconstruction semantic similarity consistently degrades when using the AdamW optimizer compared to SGD. This is consistent with our attack insight \ref{insight: optimizers}, as under the AdamW optimizer, \textit{\sysname} can only perform approximated reconstruction. For the reconstruction rate, the results show no consistent advantage between the two optimizers. In general, \textit{\sysname} is capable of reconstructing a substantial proportion ($>$55\%) of samples with high semantic fidelity, validating the effectiveness of our attack.
We further illustrate the semantic similarity distributions of all reconstructed samples across the four datasets under the Qwen2-1.5B model. From the results, we observe that using the AdamW optimizer leads to a noticeable shift in the distribution toward lower semantic similarity values compared to SGD. In particular, the reconstructed samples exhibit a unimodal high peak near the semantic similarity of 1.0 under the SGD optimizer, reflecting the \textit{exact reconstruction} nature of our attack.

\begin{figure}[t]
    \centering
    \subfigure[Rate/Bin Number]{\includegraphics[width=0.234\textwidth]{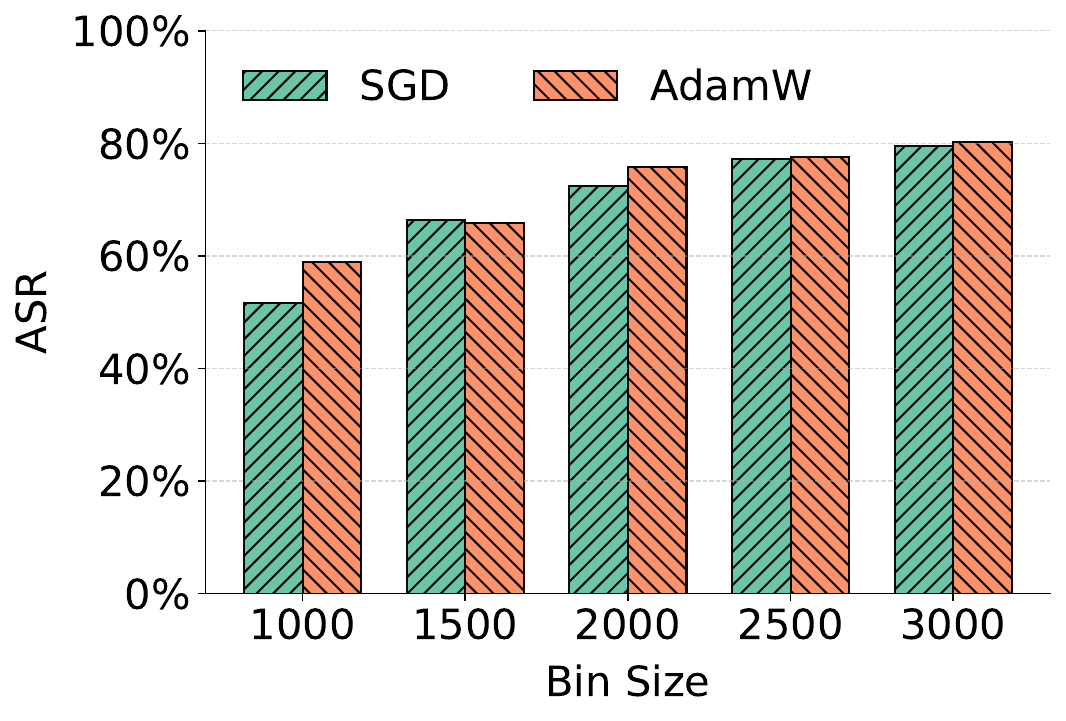}}
    \subfigure[Semantic Sim./Bin Number]{\includegraphics[width=0.234\textwidth]{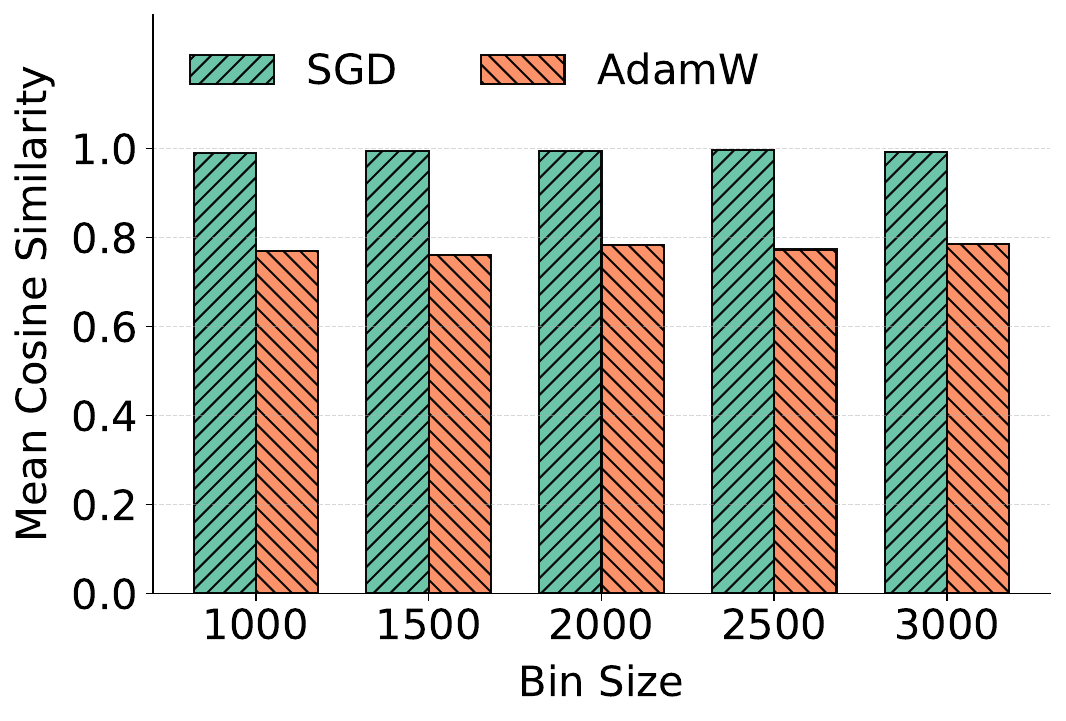}}
    \caption{\textit{\sysname} reconstruction performance across different reconstruction bin sizes under the GPT-2 model. }
    \label{fig: FL-bin size} 
\end{figure}

\textbf{Bin Size Impact:} We vary the reconstruction bin number $m$ from 1000 to 3000 to examine how the bin number $m$ correlates with the size of the target dataset $\mathrm{D}_{\mathrm{target}}$ (denoted as $d$), which is fixed as $d=800$. We conduct the experiment on the AGNews dataset and demonstrate the attack results in Fig. \ref{fig: FL-bin size}. From the results, we observe that the reconstruction rate increases when the bin size increases. This is consistent with our theoretical analysis, as more reconstruction bins reduce the probability of reconstruction conflicts and overlaps. Meanwhile, we observe that the semantic similarity remains stable across different bin sizes, indicating that the reconstruction quality is unaffected. Once reconstructed, the samples consistently exhibit high semantic fidelity.

\begin{figure}[t]
    \centering
    \subfigure[Rate/Data Size]{\includegraphics[width=0.234\textwidth]{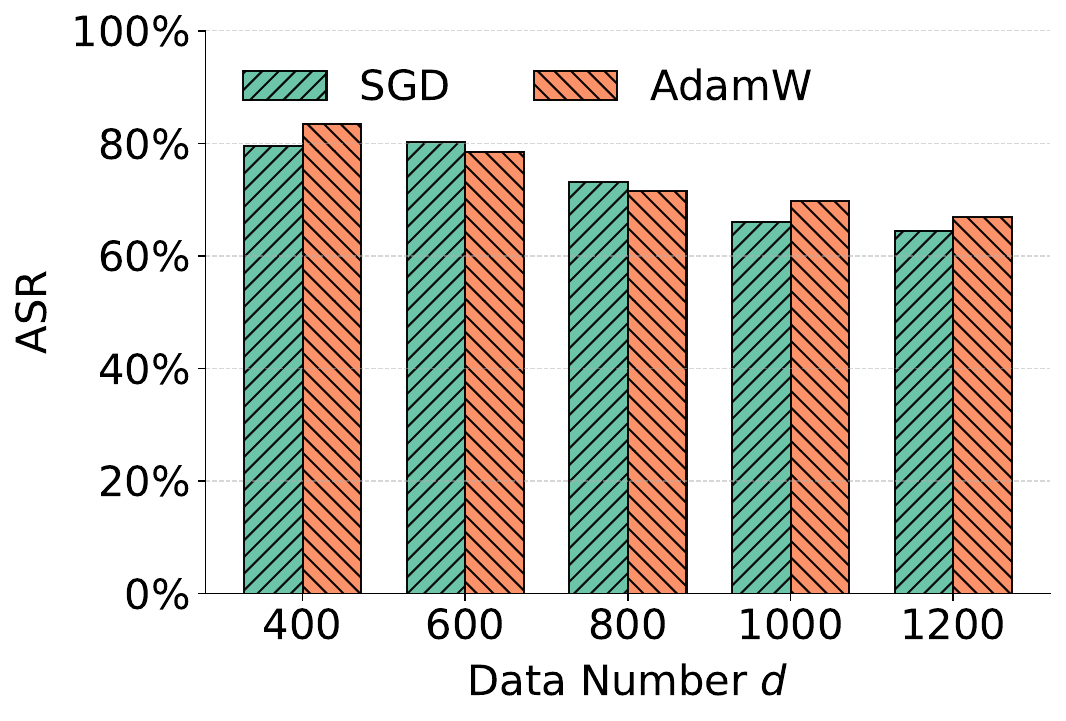}}
    \subfigure[Semantic Sim./Data Size]{\includegraphics[width=0.234\textwidth]{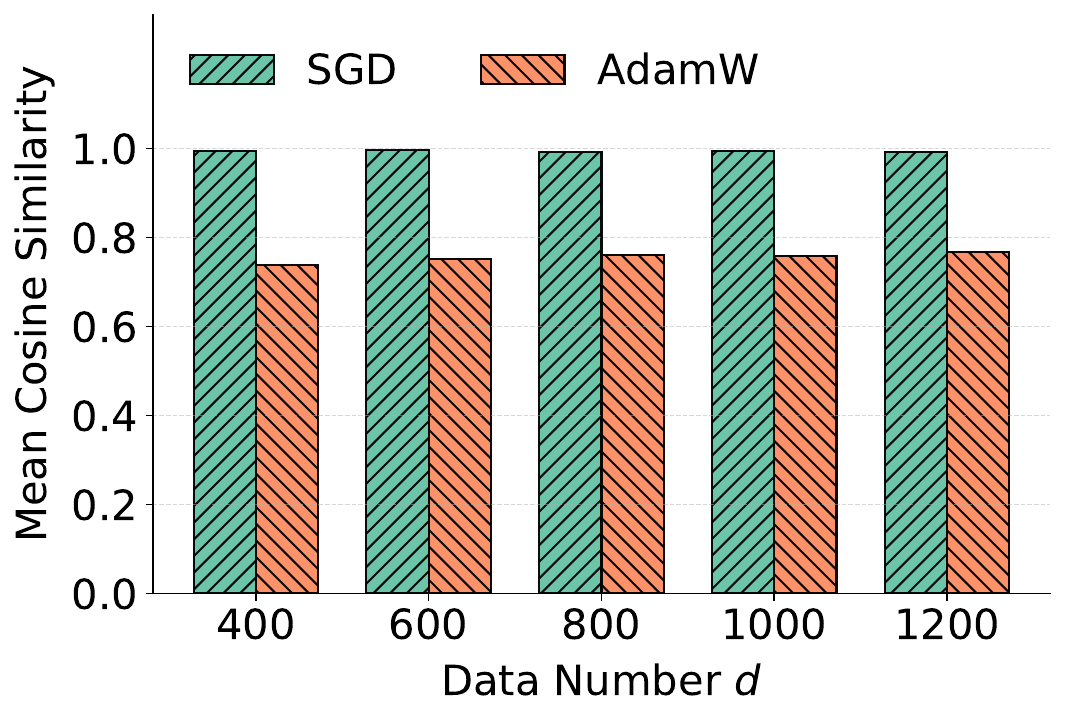}}
    \caption{\textit{\sysname} reconstruction performance across different local data sizes under the GPT-2 model. }
    \label{fig: FL-data size} 
\end{figure}

\textbf{Data Size Impact:} To further investigate the relationship between the bin number $m$ and the size $d$ of the target dataset $\mathrm{D}_{\mathrm{target}}$, we fix the bin size $m=2000$ and change $d$ from 400 to 1200. We conduct the experiment on the AGNews dataset and report the results in Fig. \ref{fig: FL-data size}. From the results, we can find that the reconstruction rates decrease when the target data size increases. However, the reconstruction semantic similarity scores are very stable across different data settings. As a conclusion for the two Sections, we find that the reconstruction rate is not decided by either the bin number $m$ or the target dataset's size $d$ alone, but the ratio between them. A larger $\frac{m}{d}$ ratio leads to better reconstruction results and vice versa. In practice, we observe that \textit{\sysname} can achieve decent attack performance when the ratio satisfies $\frac{m}{d}>2$.

\begin{figure}[t]
    \centering
    \subfigure[Rate/Non-IID Degree]{\includegraphics[width=0.234\textwidth]{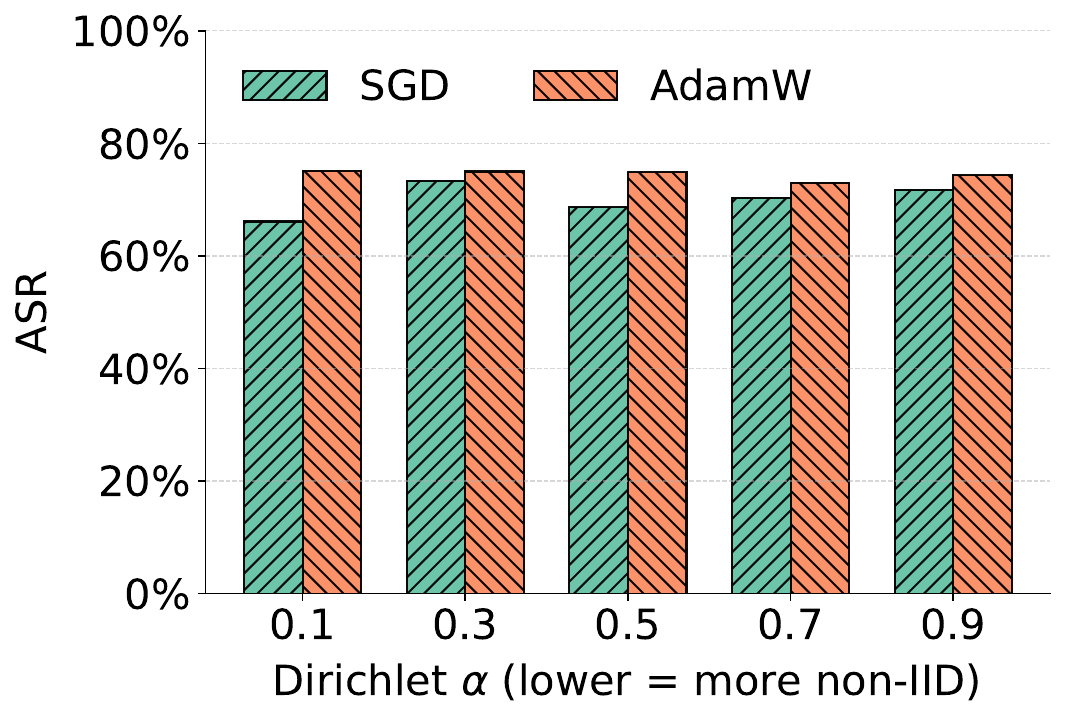}}
    \subfigure[Semantic Sim./Non-IID Degree]{\includegraphics[width=0.234\textwidth]{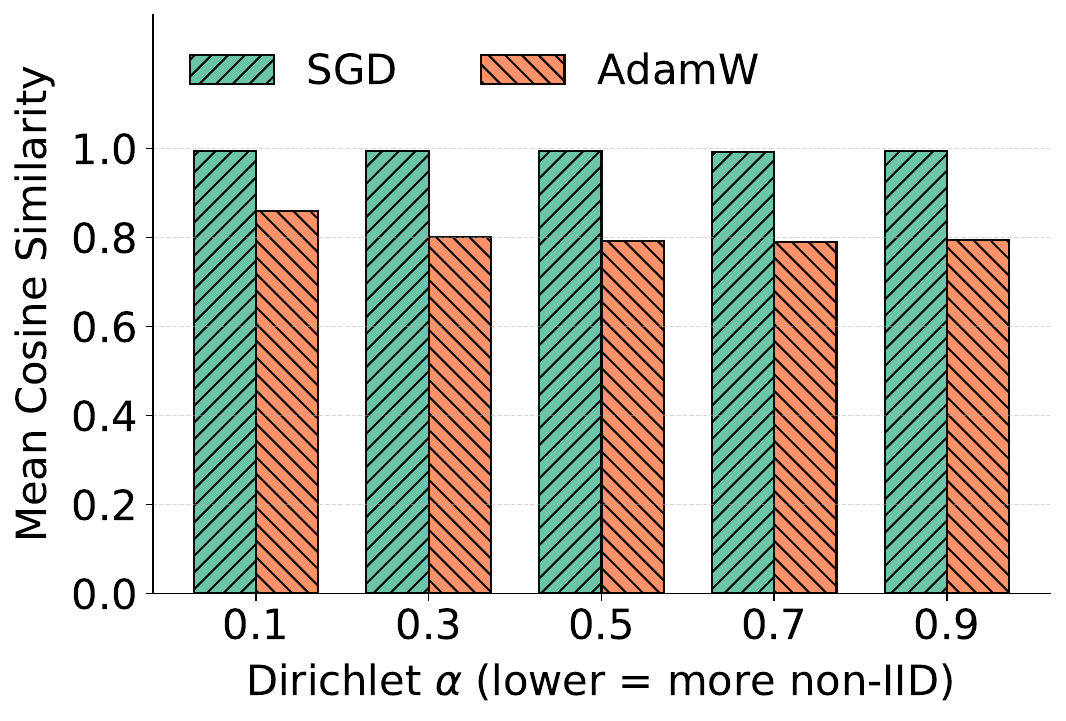}}
    \caption{\textit{\sysname} reconstruction performance across varying non-IID degrees under the GPT-2 model. }
    \label{fig: FL-noniid} 
\end{figure}

\textbf{Non-IID Impact:} We demonstrate \textit{\sysname}'s performance under different non-IID degrees (denoted by $\alpha$) in Fig. \ref{fig: FL-noniid}. We conduct our experiment on the AGNews dataset and use the well-adapted Dirichlet distribution \cite{hsu2019measuring} to partition datasets. We select non-IID degree $\alpha$ ranging from $\{0.1,0.3,0.5,0.7,0.9\}$, where a smaller $\alpha$ indicates more skewed data partition. We find that \textit{\sysname}'s performance is \textit{slightly affected} by the client's non-IID degree. In particular, the reconstruction rates are perturbed and slightly decrease when the non-IID level increases, while the reconstruction semantic similarity remains stable.
However, we clarify that this experiment is conducted under the assumption that the auxiliary dataset $D_{\mathrm{aux}}$ has a similar distribution to the overall training dataset $D_{\mathrm{train}}$, before it is partitioned to different clients. We further investigate the impact of the distribution shift between $D_{\mathrm{aux}}$ and $D_{\mathrm{train}}$ in Section 5.4.



\begin{figure}[t]
    \centering
    \subfigure[Rate/Client Number]{\includegraphics[width=0.234\textwidth]{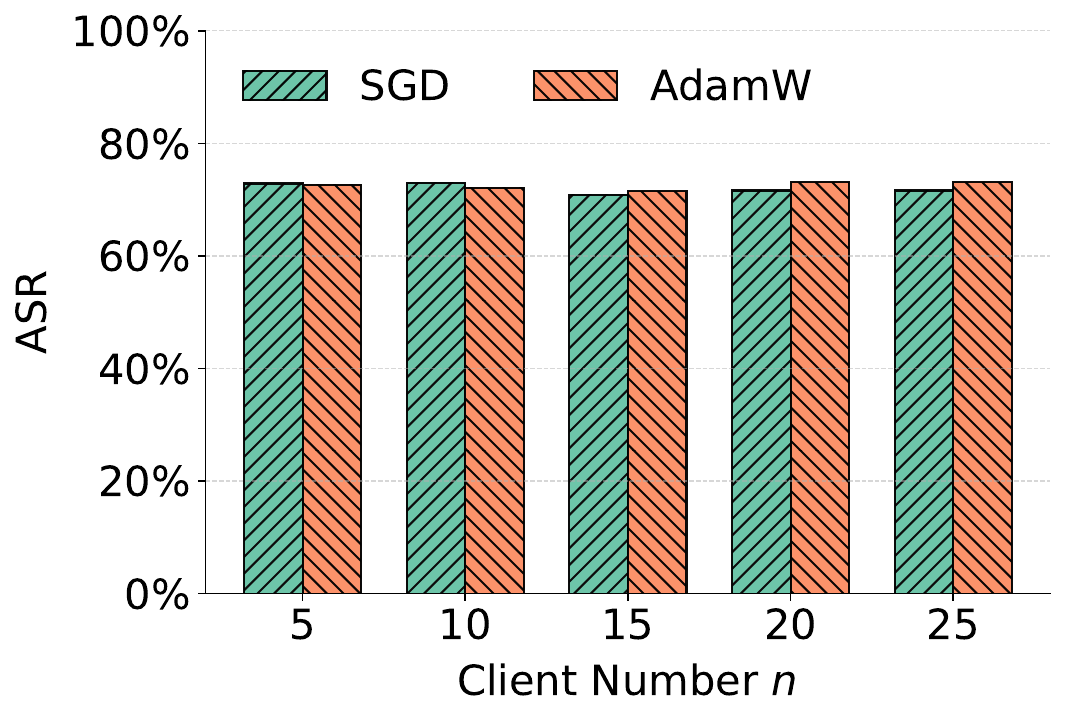}}
    \subfigure[Semantic Sim./Client Number]{\includegraphics[width=0.234\textwidth]{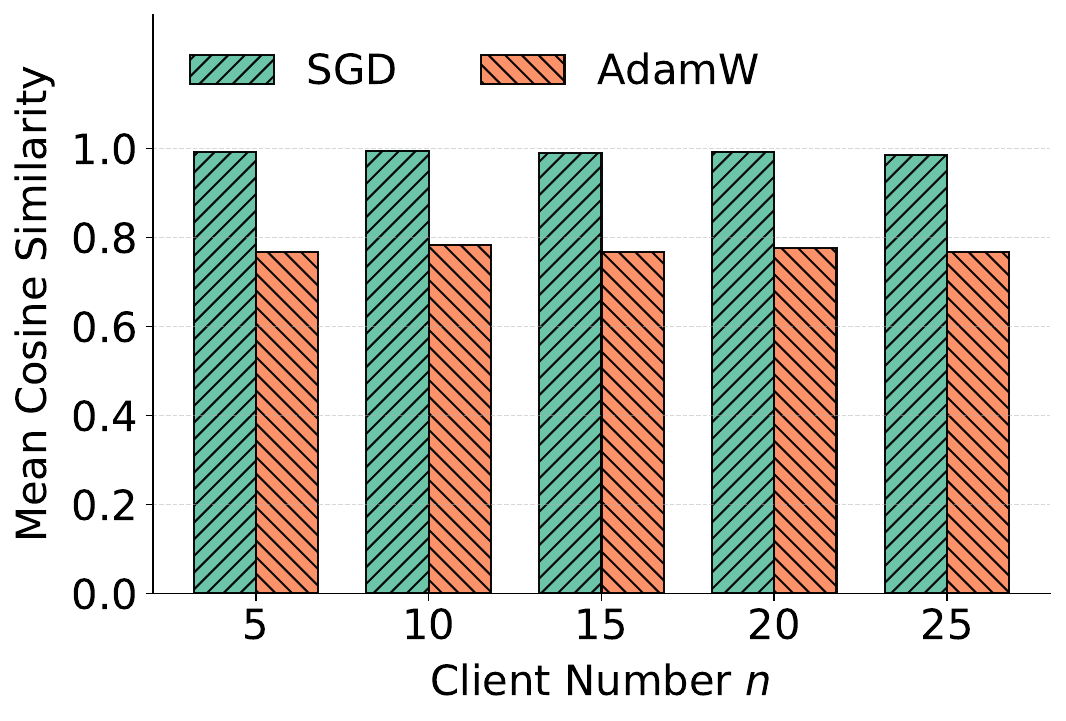}}
    \caption{\textit{\sysname} reconstruction performance across varying numbers of FL clients under the GPT-2 model. }
    \label{fig: FL-client number} 
\end{figure}

\textbf{Client Number Impact:} We evaluate \textit{\sysname}'s reconstruction performance across varying numbers of FL clients ranging from 5 to 25 on the AGNews dataset. The results are illustrated in Fig. \ref{fig: FL-client number}. We find that both the reconstruction rate and semantic similarity remain highly stable across different client numbers. This demonstrates \textit{\sysname}'s robustness in bypassing secure aggregation mechanisms, validating it can selectively reconstruct samples belonging to the victim client from a few dozen clients. 

\begin{figure*}[t]
    \centering
    \subfigure[Model Accuracy/Round]{\includegraphics[width=0.234\textwidth]{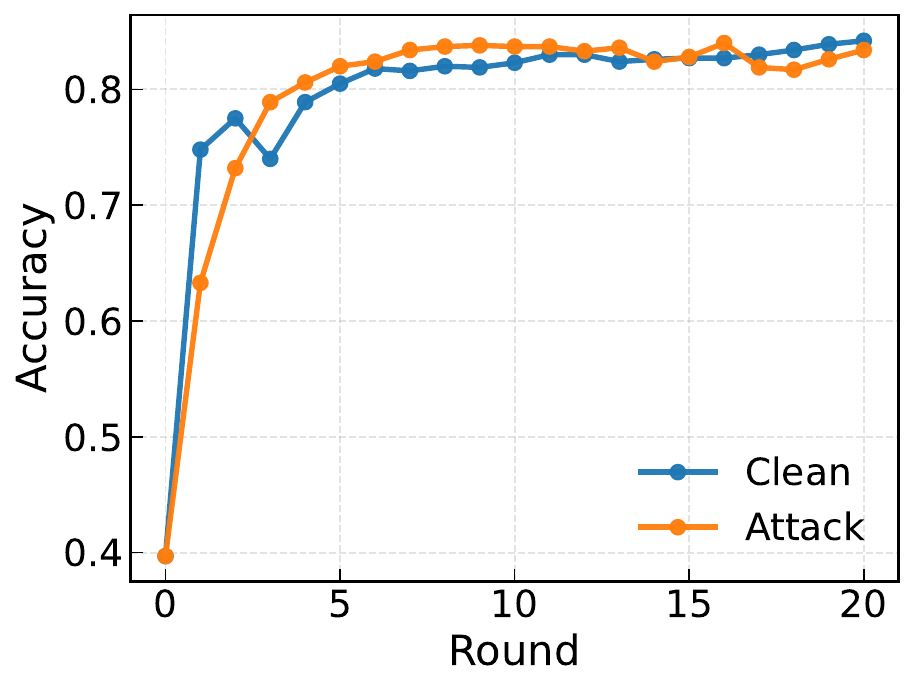}}
    \subfigure[Training Loss/Round]{\includegraphics[width=0.234\textwidth]{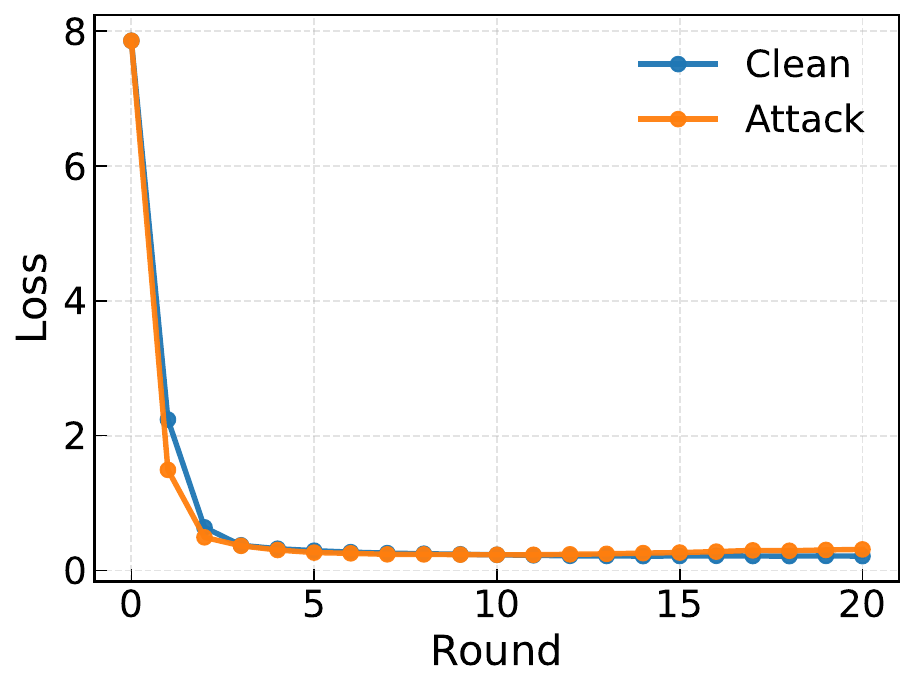}}
    \subfigure[ROC Curve/Without Attack]{\includegraphics[width=0.25\textwidth]{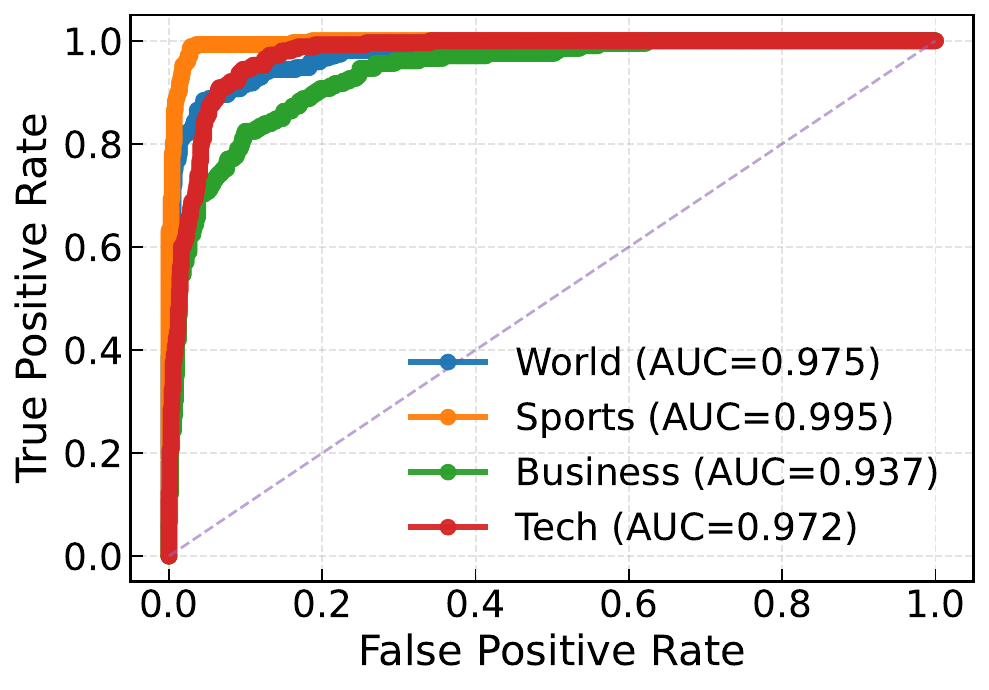}}
    \subfigure[ROC Curve/With Attack]{\includegraphics[width=0.25\textwidth]{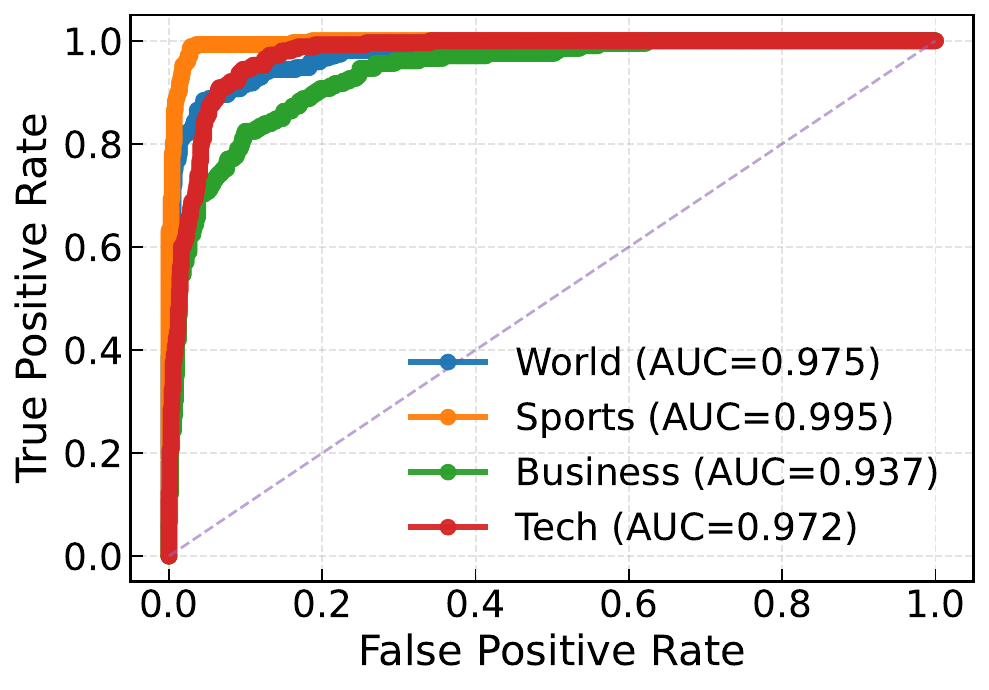}}
    \caption{GPT-2's performance with and without \textit{\sysname} on the AGNews dataset. }
    \label{fig: FL-attack-stealthiness} 
\end{figure*}

\begin{table}[t]
\centering
\caption{Model performance under different Non-IID levels (with or without attack).}
\vspace{2pt}
\small
\setlength{\tabcolsep}{4pt}
\begin{tabular}{cccccc}
\toprule
\textbf{Non-IID} & \textbf{w/ or w/o} & \textbf{Acc} & \textbf{Prec} & \textbf{Rec} & \textbf{F1} \\
\midrule
\multirow{2}{*}{0.9} 
 & w/o & 0.8530 & 0.8476 & 0.8469 & 0.8463 \\
 & w/  & 0.8420 & 0.8363 & 0.8359 & 0.8335 \\
\multirow{2}{*}{0.5} 
 & w/o & 0.8470 & 0.8413 & 0.8411 & 0.8404 \\
 & w/  & 0.8450 & 0.8379 & 0.8379 & 0.8363 \\
\multirow{2}{*}{0.1} 
 & w/o & 0.8360 & 0.8319 & 0.8287 & 0.8268 \\
 & w/  & 0.8430 & 0.8366 & 0.8359 & 0.8342 \\
\bottomrule
\end{tabular}
\label{tab: non_iid_perf}
\end{table}

\begin{table}[t]
\centering
\caption{Cross-dataset reconstruction performance on the GPT-2 model.}
\vspace{2pt}
\small
\setlength{\tabcolsep}{4pt}
\begin{tabular}{cccc}
\toprule
$\mathbf{D_{\mathrm{aux}}}\rightarrow\mathbf{D_{\mathrm{target}}}$ & \textbf{Optimizer} & \textbf{Rate} & \textbf{Sim.$\pm$Std} \\
\midrule
\multirow{2}{*}{EMRQA$\rightarrow$MedQuAD} & SGD  & 0.420 & $0.922 \pm 0.159$ \\
& AdamW   & 0.504 & $0.713 \pm 0.147$ \\
\multirow{2}{*}{MedQuAD$\rightarrow$EMRQA} & SGD & 0.525 & $0.948 \pm 0.129$ \\
& AdamW & 0.571 & $0.806 \pm 0.078$ \\
\midrule
\multirow{2}{*}{SQuAD$\rightarrow$EMRQA} & SGD   & 0.578 & $0.952 \pm 0.125$ \\
& AdamW   & 0.649 & $0.808 \pm 0.076$ \\
\multirow{2}{*}{EMRQA$\rightarrow$SQuAD} & SGD & 0.672 & $0.978 \pm 0.094$ \\
& AdamW & 0.735 & $0.749 \pm 0.100$ \\
\midrule
\multirow{2}{*}{SQuAD$\rightarrow$MedQuAD} & SGD & 0.245 & $0.969 \pm 0.118$ \\
& AdamW & 0.469 & $0.723 \pm 0.144$ \\
\multirow{2}{*}{MedQuAD$\rightarrow$SQuAD} & SGD & 0.537 & $0.978 \pm 0.094$ \\
& AdamW & 0.562 & $0.738 \pm 0.109$ \\
\bottomrule
\end{tabular}
\label{tab: cross-dataset-attack}
\end{table}

\subsection{Attack Stealthiness}  

In Section 4.5, we provide a theoretical proof for why the privacy backdoor introduces no model performance impact. In this section, we further provide an experimental validation for this attack property. We conduct the FL training for 20 rounds under the default FL setting for the GPT-2 model on the AGNews dataset with and without \textit{\sysname} attack. We report the model accuracy, training performance versus training rounds, as well as the per-class ROC curves in Fig. \ref{fig: FL-attack-stealthiness}. From the results, we can find that the model performance is almost identical with and without \textit{\sysname} attack. In addition, we report the model's accuracy, precision, recall, and F1 scores on different non-IID levels with and without attack in Tab. \ref{tab: non_iid_perf}. From the results, we observe only minor differences in model performance across all evaluation metrics with or without our attack. This further highlights that our privacy backdoor introduces no noticeable performance degradation. We regard this as a crucial property for preserving attack stealthiness, as clients observe no detectable performance differences when the attack is launched. In addition, we find that the overall model training performance is not significantly affected by the non-IID degree. This may be due to GPT-2’s powerful modeling ability and the simplicity of the classification task.


\begin{table*}[t]
\centering
\small
\setlength{\tabcolsep}{4pt}
\caption{\textit{\sysname}'s reconstruction performance under LoRA fine-tuning across different models and datasets.}
\vspace{2pt}
\begin{tabular}{lccccccc}
\toprule
\multirow{2}{*}{\textbf{Dataset}} &
\multirow{2}{*}{\textbf{Token Len.}} &
\multirow{2}{*}{\textbf{Optimizer}} &
\multicolumn{1}{c}{\textbf{BERT}} &
\multicolumn{1}{c}{\textbf{GPT-2}} &
\multicolumn{1}{c}{\textbf{Qwen2-1.5B}} &
\multicolumn{1}{c}{\textbf{Llama3-3B}} \\
\cmidrule(r){4-7}
& & & Rate$|$Sim.$\pm$Std & Rate$|$Sim.$\pm$Std & Rate$|$Sim.$\pm$Std & Rate$|$Sim.$\pm$Std \\
\midrule
\multirow{2}{*}{AGNews}
 & \multirow{2}{*}{128}
 & SGD   & 0.682$|$0.995$\pm$0.022 & 0.708$|$0.993$\pm$0.047 & 0.688$|$0.998$\pm$0.025 & 0.719$|$0.998$\pm$0.025 \\
 &       & AdamW & 0.734$|$0.773$\pm$0.087 & 0.744$|$0.729$\pm$0.120 & 0.740$|$0.820$\pm$0.104 & 0.759$|$0.918$\pm$0.071 \\
\midrule
\multirow{2}{*}{SQuAD}
 & \multirow{2}{*}{384}
 & SGD   & 0.680$|$0.988$\pm$0.060 & 0.657$|$0.971$\pm$0.110 & 0.733$|$0.978$\pm$0.100 & 0.711$|$0.994$\pm$0.045 \\
 &       & AdamW & 0.719$|$0.714$\pm$0.097 & 0.749$|$0.719$\pm$0.099 & 0.744$|$0.671$\pm$0.109 & 0.733$|$0.783$\pm$0.089 \\
\midrule
\multirow{2}{*}{mSQuAD}
 & \multirow{2}{*}{384}
 & SGD   & 0.796$|$0.955$\pm$0.119 & 0.686$|$0.971$\pm$0.096 & 0.719$|$0.972$\pm$0.100 & 0.723$|$0.966$\pm$0.106 \\
 &       & AdamW & 0.735$|$0.801$\pm$0.078 & 0.710$|$0.810$\pm$0.075 & 0.700$|$0.752$\pm$0.072 & 0.706$|$0.829$\pm$0.067 \\
\midrule
\multirow{2}{*}{GSM8K}
 & \multirow{2}{*}{512}
 & SGD   & -- & 0.630$|$0.977$\pm$0.099 & 0.762$|$0.965$\pm$0.129 & 0.741$|$0.974$\pm$0.112 \\
 &       & AdamW & -- & 0.626$|$0.504$\pm$0.074 & 0.762$|$0.683$\pm$0.101 & 0.787$|$0.864$\pm$0.065 \\
\bottomrule
\end{tabular}
\\[2mm]
\footnotesize{BERT is excluded from GSM8K because it is an encoder-only model for classification/span extraction.}
\label{tab: lora-perf}
\end{table*}

\subsection{Cross-Dataset Transferability}

In this section, we investigate \textit{\sysname}'s cross-dataset attack transferability, where the adversary possesses an auxiliary dataset $D_{\mathrm{aux}}$ that differs from the target dataset $D_{\mathrm{target}}$. This setting reflects a more realistic attack scenario, as an adversary is unlikely to obtain an auxiliary dataset that perfectly matches the target distribution. Instead, the attacker may only have broad domain knowledge and access to an in-domain dataset with \textit{similar characteristics}.

For the experiment, we adopt the default FL settings and introduce an additional dataset, named MedQuAD \cite{ben2019question}. This is a large-scale medical question-answering dataset automatically collected from the U.S. National Institutes of Health (NIH) webpages, containing over 47000 consumer-style question–answer pairs covering diseases, drugs, symptoms, and treatments. Unlike EMRQA-mSQuAD, which contains longer and clinically detailed EMR text, MedQuAD features shorter, consumer-oriented questions and concise public health answers. Besides MedQuAD, we also include the EMRQA-mSQuAD and SQuAD datasets in our experiments, as all three share the same question–answer format suitable for consistent evaluation. In each case, we designate one dataset as the auxiliary dataset and another as the reconstruction target, then swap their roles to assess bidirectional transferability. This setup is feasible under the assumption that the adversary possesses basic knowledge of the FL local training data and overall task. We demonstrate the attack performance in Tab. \ref{tab: cross-dataset-attack}. 

The first pair of datasets, EMRQA-mSQuAD and MedQuAD, are both biomedical datasets and represent the attack scenario in which the adversary possesses an in-domain auxiliary dataset. We use them to show that \textit{\sysname} transfers well across similar biomedical sources, achieving consistently strong reconstruction performance. The second pair (SQuAD and EMRQA-mSQuAD) and the third pair (SQuAD and MedQuAD) emulate the scenario where the attacker only has a general auxiliary dataset that is not necessarily from the target domain. However, despite this mismatch, our attack still reconstructs a substantial portion of the samples. These results indicate that different QA datasets with the same format share rich structural and semantic commonalities that \textit{\sysname} can exploit. We further observe that when the auxiliary dataset is more domain-specific, such as EMRQA-mSQuAD or MedQuAD, the attack transfers more effectively than when using a general-domain dataset like SQuAD. This improvement is likely attributed to the reduced domain mismatch between the auxiliary and target datasets.

\subsection{Attack under LoRA}

We evaluate \textit{\sysname}'s performance when the transformer blocks are under LoRA fine-tuning. We adopt the default FL setting and conduct experiments across all four datasets. For the LoRA parameters, we use a rank of 8, a scaling factor of 32, and a dropout rate of 0.1, applied to the attention and projection layers within the transformer blocks for efficient adaptation. However, we note that the embedding block is still using a parallel adapter to enable the attack. We conduct the experiments on all four datasets and evaluate both the reconstruction rate and semantic similarity.

We report the attack results in Table \ref{tab: lora-perf}. The data show that \textit{\sysname}’s effectiveness is essentially unchanged under LoRA fine-tuning and closely matches the results in Table \ref{tab: general-performance}. In some settings, the attack even performs slightly better under LoRA. This robustness arises because \textit{\sysname} only relies on the gradients of the memorization layer to execute the attack and is agnostic to whether the subsequent layers are trained with LoRA, parallel adapters, or even full fine-tuning, as long as the gradients are properly backpropagated into the backdoor. These findings demonstrate that our attack can be applied to LoRA fine-tuning settings. We further provide attack performance under LoRA for larger Qwen2-7B and Llama3-8B in Appendix E.


\section{Discussion}
\label{sec: discussion}

\textbf{Potential Defense: } \sysname is designed as a stealthy privacy attack that embeds a backdoor by crafting a seemingly benign PEFT adapter while preserving model utility. To defend against this attack, a defender may leverage parameter-level artifacts, such as repeated row vectors or nonstandard bias, to detect the presence of possible backdoor. Even though the attacker may still perform an adaptive attack to evade detection, it imposes a constrain on the design and would lead to decreased effectiveness. 


\noindent\textbf{Scalability \& Memory Space Trade-off: }
As we have discussed in Section \ref{sec: method}, \sysname’s reconstruction scalability is largely determined by the number of neurons $m$ in the memorization layer. Since \sysname assigns one neuron per sample, the memory overhead grows approximately linearly with $m$. In practice, reconstructing 2000 samples requires about 200MB for Llama-3.2, which is less than 3\% of the model size. However, the memorization layer cannot be scaled indefinitely, and there is a clear scalability–memory trade-off.
To accommodate larger target sets under a fixed memory budget, the adversary can form reconstruction bins using only a subset of $\psi$, thereby reconstructing only the samples mapped to that subset and reducing the required neurons. Scalability can be further improved by repeating the attack across multiple FL rounds, increasing the number of reconstruction trials, and cumulatively recovering more samples. Furthermore, stronger prior information for the domain-specific fine-tuning process also allows a more compact representation of key features in samples.

\section{Conclusion}

In this paper, we propose \sysname, a practical data reconstruction attack against federated language model fine-tuning. The attack is realized by crafting a seemingly benign PEFT adapter attached to the embedding block into a privacy backdoor that memorizes per-sample model updates at distinct neurons during fine-tuning, which enables closed-form reconstruction of private client data from them. \sysname is effective under realistic finetuning settings, achieving high-fidelity reconstruction for large training batches and long sequences, and remains effective under stateful optimizers. Extensive experiments across multiple model families and datasets demonstrate the effectiveness and generality of \sysname. Our findings expose a critical privacy risk in current federated fine-tuning pipelines and highlight the need for stronger safeguards against this emerging privacy attack.


\clearpage
\appendix
\section*{Ethical Considerations}

We conduct a stakeholder-based ethics analysis and distinguish impacts arising from the \emph{research process} (data handling and experimentation) and the \emph{publication of results} (deployment and downstream use). 

\vspace{3pt}\noindent\textbf{Stakeholders.}
Our work may affect: (1) \textit{Federated learning clients/data owners}, whose private fine-tuning data may be exposed if the vulnerability is exploited; (2) \textit{data subjects} represented in client datasets, who may experience privacy harm if their information is reconstructed; (3) \textit{system developers and deployers}, who may incur mitigation, compliance, and incident-response costs; and (4) \textit{the research community (including our team)}, which benefits from clearer threat understanding but also bears responsibility to minimize dual-use risk and ensure methodological rigor.

\vspace{3pt}\noindent\textbf{Impacts.}
\textit{Research process (data handling):}
We evaluate \sysname in a \emph{controlled setting} and do not probe real-world deployments or access non-consensual private user data. Any reconstructed outputs are treated as sensitive artifacts and handled to minimize exposure, supporting \emph{Respect for Persons} and
\emph{Respect for Law and Public Interest}.
\textit{Publication (deployment and application):}
Publication has both benefits and risks. On the positive side, our results clarify a realistic privacy vulnerability in federated parameter-efficient fine-tuning and inform safer system design. On the negative side, disclosure may enable misuse if mitigations are not adopted, potentially increasing privacy risk.

\vspace{3pt}\noindent\textbf{Potential Harms.}
\textit{Tangible harms} include leakage of sensitive or proprietary text, with possible reputational or financial harm to clients and operational harm to deployers.
\textit{Rights-based harms} include violations of privacy expectations for data owners and data subjects, even without immediate measurable loss.

\vspace{3pt}\noindent\textbf{Mitigations.}
\textit{Implemented mitigations:}
We restrict our evaluation to controlled experiments and treat reconstructed outputs as sensitive artifacts, minimizing unnecessary exposure.
\textit{Recommended mitigations:}
We recommend adapter provenance checks and pre-deployment auditing, especially for adapters obtained from untrusted sources. In our analysis, \sysname leaves parameter-level artifacts---e.g., repeated row vectors and nonstandard bias initialization in the memorization and output layers---that deviate from typical adapter initialization and training patterns. This motivates practical screening for backdoor-like constructions, alongside complementary integrity checks and monitoring.

\vspace{3pt}\noindent\textbf{Justification for Research.}
We conduct and publish this study to enable timely detection and mitigation of a plausible privacy vulnerability in federated parameter-efficient fine-tuning. Under \emph{Beneficence}, the expected defensive value of exposing this vulnerability and providing mitigation guidance outweighs residual dual-use risk. Under \emph{Respect for Persons}, we avoid violating individuals' rights by not using non-consensual real-user data and by constraining our evaluation to a controlled setting with careful handling of reconstruction outputs. Finally, responsible disclosure serves the public interest by helping deployers update threat models and adopt practical safeguards, supporting \emph{Justice} and \emph{Respect for Law and Public Interest}.

\section*{Open Science}
Our code is available at: \url{https://github.com/shishishi123/NeuroImprint/}. 

\bibliographystyle{plain}
\bibliography{reference}

\clearpage
\appendix

\section*{Appendix A.~Adam Optimizer}
\label{Adam optimizer}

The Adam optimizer is an adaptive stochastic optimization algorithm that maintains exponential moving averages of both the first- and second-order moments of the gradients \cite{kingma2014adam}. It has been widely used as the de facto training optimizer for complex machine learning optimization tasks. 
For the federated fine-tuning setting, each local client performs local training with it for $T$ rounds. At iteration $t\in\{1,2,3,\cdots,T\}$, given gradient $g_t = \nabla_{\theta} \mathcal{L}(\theta_t)$, Adam optimizer updates parameters as:
\begin{equation}
\begin{aligned}
m_t &= \beta_1 m_{t-1} + (1 - \beta_1) g_t, \\
v_t &= \beta_2 v_{t-1} + (1 - \beta_2) g_t^2, \\
\hat{m}_t &= \frac{m_t}{1 - \beta_1^t}, \quad 
\hat{v}_t = \frac{v_t}{1 - \beta_2^t}, \\
\theta_{t+1} &= \theta_t - \alpha \frac{\hat{m}_t}{\sqrt{\hat{v}_t} + \epsilon}.
\end{aligned}
\end{equation}
Here, $\alpha$ is the learning rate, $\beta_1$ and $\beta_2$ are exponential decay rates (typically 0.9 and 0.999), and $\epsilon$ is a small constant for numerical stability. We can clearly observe that the raw gradients $g_t$ are distorted during the training process. The AdamW optimizer is a decoupled variant of Adam that applies weight decay directly to the parameters rather than through the gradients, leading to more stable optimization and better generalization \cite{loshchilov2019decoupled}.

\section*{Appendix B.~Single-Step Reversion}

At the first step ($m_0=0, v_0=0$), the model update under the Adam optimizer can be approximated as follows:
\begin{equation}
\begin{aligned}
m_1 &=(1 - \beta_1) g_1, \quad
v_1 =(1 - \beta_2) g_1^2, \\
\hat{m}_1 &= \frac{m_t}{1 - \beta_1}=g_1, \quad 
\hat{v}_1 = \frac{v_t}{1 - \beta_2}=g_1^2, \\
\theta_{1} &= \theta_0 - \alpha \frac{\hat{m}_1}{\sqrt{\hat{v}_1} + \epsilon}.
\end{aligned}
\end{equation}
Therefore, $\Delta \theta_1=\alpha \frac{\hat{m}_1}{\sqrt{\hat{v}_1} + \epsilon}=\alpha\frac{g_1}{g_1+\epsilon}\approx \alpha \mathrm{sign}(g_1)$.

\section*{Appendix C.~Experiment Datasets}
The AG News dataset is a topic-classification dataset built from news headlines and descriptions collected via the AG aggregator, with 120,000 training and 7,600 test samples evenly split across four classes (World, Sports, Business, Sci/Tech). The Stanford Question Answering Dataset (SQuAD v1.1) is a Wikipedia-based extractive QA benchmark comprising a context paragraph, a question, and an answer span that appears verbatim in the context. The dataset consists of 87,599 training and 10,570 testing examples. EMRQA-mSQuAD dataset remaps EMRQA into the SQuAD v1.1 schema, pairing biomedical passages with targeted questions and short extractive spans. It contains approximately 12k span-based QA pairs, and we treat it as SQuAD-style extractive QA with the same formatting. GSM8K is a curated set of grade-school math word problems for multi-step reasoning, with 7,473 training and 1,319 test problems. The samples are in the format of Question → Final number.

\section*{Appendix D.~Reconstruction Examples}

We provide reconstruction examples for the AGNews, GSM8K, and EMRQA-mSQuAD datasets in Fig. \ref{fig: reconstruction-examples-appendix}. For each dataset, we show the original text, reconstructions under SGD and AdamW, and the LLM-refined results.

\clearpage

\begin{figure*}[h]
  \centering
  \begin{tcolorbox}[colback=gray!15, colframe=gray!40, boxrule=0.4pt, width=\linewidth, left=6pt, right=6pt, top=4pt, bottom=4pt, sharp corners, enhanced]
    \textbf{Original Text (AGNews)}:\\[2pt]
    \textbf{News:} Non-OPEC Nations Should Up Output-Purnomo JAKARTA (Reuters) - Non-OPEC oil exporters should consider increasing output to cool record crude prices, OPEC President Purnomo Yusgiantoro said on Sunday.\\
    \textbf{Category:} Business

    \vspace{2pt}
    \textbf{Reconstructed Text 1 (SGD)}:\\[2pt]
    \textbf{News:} \textcolor{blue}{Non-OPEC Nations Should Up Output-Purnomo JAKARTA (Reuters) - Non-OPEC oil exporters should consider increasing output to cool record crude prices, OPEC President Purnomo Yusgiantoro said on Sunday.}\\
    \textbf{Category:} \textcolor{blue}{Business}

    \vspace{2pt}
    \textbf{Reconstructed Text 2 (AdamW)}:\\[2pt]
    \textbf{News:} {\color{red}Non-OPEC Nations Should Up Output-Purnomo} \textless{} JAKARTA ((Reuters)) ! {\color{red}Non-OPEC oil exporters should consider} \textless{} {\color{red}increasing output} onto {\color{red}cool record crude prices, OPEC President} \textless{} nurnomo {\color{red}Yusgiantoro said} onto {\color{red}Sunday.}\\
    \textbf{Category:} \textless{} {\color{red}Business}

    \vspace{2pt}
    \textbf{LLM Rephrased (AdamW)}:\\[2pt]
    \textbf{Context:} \textcolor{brown}{JAKARTA (Reuters) — Non-OPEC oil exporters should consider increasing production to help cool record-high crude prices, OPEC President Purnomo Yusgiantoro said on Sunday.}\\
    \textbf{Category:} {\color{brown}Business}
    
    \vspace{6pt}
    \textbf{Original Text (GSM8K)}:\\[2pt]
    \textbf{Problem:} Sarah bought 4 packs of gum. Each pack contains 12 sticks. She gave 6 sticks to her friend and chewed 3 herself. How many sticks of gum does she have left?\\
    \textbf{Answer:} 39

    \vspace{2pt}
    \textbf{Reconstructed Text 1 (SGD)}:\\[2pt]
    \textbf{Problem:} \textcolor{blue}{Sarah bought 4 packs of gum. Each pack contains 12 sticks. She gave 6 sticks to her friend and chewed 3 herself. How many sticks of gum does she have left?}\\
    \textbf{Answer:} \textcolor{blue}{39}

    \vspace{2pt}
    \textbf{Reconstructed Text 2 (AdamW)}:\\[2pt]
    \textbf{Problem:} \textcolor{red}{Sarah} purchased  \textcolor{red}{four gum packs} with \textcolor{red}{12 sticks} each, \textcolor{red}{gave six} to a \textcolor{red}{friend} and \textcolor{red}{chewed three}. \textcolor{red}{How many sticks of gum} remain?\\
    \textbf{Answer:} {\color{red}39}

    \vspace{2pt}
    \textbf{LLM Rephrased (AdamW)}:\\[2pt]
    \textbf{Problem:} \textcolor{brown}{Sarah purchased four packs of gum with 12 sticks each, gave six sticks to a friend, and chewed three herself. How many sticks of gum remain?}\\
    \textbf{Answer:} {\color{brown}39}

    \vspace{6pt}
    \textbf{Original Text (mSQUAD)}:\\ [2pt] 
    \textbf{Content:} This 70-year-old female with CHF, coronary artery disease, diabetes, peripheral vascular disease, and chronic renal insufficiency was admitted on 0/5/06 for weakness and confusion. Her hospital course was complicated by worsening cardiac function with minimal improvement on milrinone and decreasing urine output despite diuretics and also gross gastrointestinal bleeding with melanotic stool while she was on Coumadin for atrial fibrillation. In addition, there was concern for sepsis and she was placed on antibiotics with levofloxacin, Flagyl, and vancomycin. She required a transfer to the Cardiac Care Unit on 9/15/06 for further medical therapy for poor cardiac output, a possible need for CVVH, given volume overload in the setting of renal failure, and work-up of GIB. Her code status was DNR/DNI, but was changed to comfort measures only on 1/17/06 due to a large ascending colorectal mass with ulcerations. Being CMO status, she was removed of all pressors and antibiotics and made comfortable sedated on fentanyl and Versed. She was then extubated for comfort with family present and had agonal breathing with episodes of apnea and was given additional sedation for comfort. The patient drew her last breath at 2:20 p.m. with family present and was pronounced dead at 2:20 p.m. on 1/17/06. Family declined autopsy.\\
    \textbf{Question:} What does the patient take antibiotics for \\
    \textbf{Answer:} fibrillation. In addition, there was concern for sepsis
  \end{tcolorbox}
\end{figure*}

\clearpage
    
\begin{figure*}[h] \centering \begin{tcolorbox}[colback=gray!15, colframe=gray!40, boxrule=0.4pt, width=\linewidth, left=6pt, right=6pt, top=4pt, bottom=4pt, sharp corners, enhanced]
    \textbf{Reconstructed Text 1 (SGD)}:\\ [2pt] 
    \textbf{Context:} \textcolor{blue}{This 70-year-old female with CHF, coronary artery disease, diabetes, peripheral vascular disease, and chronic renal insufficiency was admitted on 0/5/06 for weakness and confusion. Her hospital course was complicated by worsening cardiac function with minimal improvement on milrinone and decreasing urine output despite diuretics and also gross gastrointestinal bleeding with melanotic stool while she was on Coumadin for atrial fibrillation. In addition, there was concern for sepsis and she was placed on antibiotics with levofloxacin, Flagyl, and vancomycin. She required a transfer to the Cardiac Care Unit on 9/15/06 for further medical therapy for poor cardiac output, a possible need for CVVH, given volume overload in the setting of renal failure, and work-up of GIB. Her code status was DNR/DNI, but was changed to comfort measures only on 1/17/06 due to a large ascending colorectal mass with ulcerations. Being CMO status, she was removed of all pressors and antibiotics and made comfortable sedated on fentanyl and Versed. She was then extubated for comfort with family present and had agonal breathing with episodes of apnea and was given additional sedation for comfort. The patient drew her last breath at 2:20 p.m. with family present and was pronounced dead at 2:20 p.m. on 1/17/06. Family declined autopsy.}\\
    \textbf{Question:} \textcolor{blue}{What does the patient take antibiotics for} \\
    \textbf{Answer:} \textcolor{blue}{fibrillation. In addition, there was concern for sepsis}

    \vspace{6pt}
    \textbf{Reconstructed Text 2 (Adamw)}: \\ [2pt]
    \textbf{Context:} There \textcolor{red}{70}." \textcolor{red}{year}"." \textcolor{red}{old} \textcolor{red}{female} toward \textcolor{red}{CHF}., \textcolor{red}{coronary artery disease}., \textcolor{red}{diabetes}., \textcolor{red}{peripheral vascular disease}., argues \textcolor{red}{chronic renal insufficiency} seemed \textcolor{red}{admitted} onto \textcolor{red}{0}.'\textcolor{red}{5}.'\textcolor{red}{06}., \textcolor{red}{weakness} argues \textcolor{red}{confusion}., \textcolor{red}{Her hospital course} seemed \textcolor{red}{complicated}], \textcolor{red}{worsening cardiac function} toward \textcolor{red}{minimal improvement} onto \textcolor{red}{milrinone} argues \textcolor{red}{decreasing urine output} despite \textcolor{red}{diuretics} argues maintains \textcolor{red}{gross gastrointestinal bleeding} toward \textcolor{red}{melanotic stool} while herself seemed onto \textcolor{red}{Coumadon}da.,],rial \textcolor{red}{fibrillation}., Its \textcolor{red}{addition}., someone seemed \textcolor{red}{concern}., \textcolor{red}{sepsis} argues herself seemed \textcolor{red}{placed} onto \textcolor{red}{antibiotics} toward \textcolor{red}{levofloxaconda}., \textcolor{red}{Flagyl}., argues \textcolor{red}{vancomycin}., \textcolor{red}{She required} several \textcolor{red}{transfer}., any \textcolor{red}{Cardiac Care Unit} onto \textcolor{red}{9}.'\textcolor{red}{15}.'\textcolor{red}{06}., further \textcolor{red}{medical therapy}., \textcolor{red}{poor cardiac output}., several \textcolor{red}{possible need}., \textcolor{red}{CVVH}a., given \textcolor{red}{volume overload} according any setting., \textcolor{red}{renal failure}., argues work."up., \textcolor{red}{GIB}., \textcolor{red}{Her code status} seemed \textcolor{red}{DNR}.'\textcolor{red}{DNI}., something seemed changed., \textcolor{red}{comfort measures only} onto 60.'\textcolor{red}{17}.'\textcolor{red}{06} due., several sizable \textcolor{red}{ascending colorectal mass} toward \textcolor{red}{ulcerations}., Being \textcolor{red}{CMO} status., herself seemed removed., every \textcolor{red}{pressors} argues \textcolor{red}{antibiotics} argues kept \textcolor{red}{comfortable} sedated onto \textcolor{red}{fentanyl} argues \textcolor{red}{Versed}., She seemed themselves \textcolor{red}{extubated}., \textcolor{red}{comfort} toward \textcolor{red}{family} occasion argues happened \textcolor{red}{agonal breathing} toward \textcolor{red}{episodes of apnea} argues seemed given \textcolor{red}{additional sedation}., \textcolor{red}{comfort}., Their \textcolor{red}{patient} drew her \textcolor{red}{last breath}], 72,'20 str.,m., toward \textcolor{red}{family} occasion argues seemed \textcolor{red}{pronounced dead}], 72,'20 str.,m., onto 60.'17.'06., \textcolor{red}{Family declined autopsy.}, \\
    \textbf{Question:} 'These doesn any \textcolor{red}{patient stay antibiotics.},  \\
    \textbf{Answer:}
    \textcolor{red}{fibrillation. In addition, there was concern for sepsis} 

    \vspace{6pt} 
    \textbf{LLM Rephrased (AdamW)}:\\ [2pt] 
    \textbf{Context:} \textcolor{brown}{A 70-year-old female with a history of congestive heart failure, coronary artery disease, diabetes, peripheral vascular disease, and chronic renal insufficiency was admitted on 05/05/06 with weakness and confusion. Her hospital course was complicated by worsening cardiac function with minimal improvement on milrinone, decreasing urine output despite diuretics, and gross gastrointestinal bleeding with melanotic stool while on Coumadin for atrial fibrillation. In addition, there was concern for sepsis, and she was started on antibiotics including levofloxacin, Flagyl, and vancomycin. She required several transfers, including to the Cardiac Care Unit on 09/15/06, for further management of poor cardiac output and possible need for CVVH due to renal failure and volume overload. Workup also revealed gastrointestinal bleeding. Her code status was DNR/DNI, later changed to comfort measures only on 10/17/06 due to a large ascending colorectal mass with ulcerations. Following transition to CMO status, pressors and antibiotics were discontinued, and she was kept comfortable and sedated on fentanyl and Versed. She was extubated for comfort in the presence of family and developed agonal breathing and episodes of apnea, for which she received additional sedation. The patient passed away at 7:20 a.m. on 10/17/06. The family declined autopsy. }\\
    \textbf{Question:} \textcolor{brown}{Did the patient stay on antibiotics?} \\
    \textbf{Answer:} \textcolor{brown}{The patient was treated for sepsis with antibiotics, including levofloxacin, Flagyl, and vancomycin.}
    \end{tcolorbox} 
    \caption{\textit{\sysname} reconstruction examples on the AGNews, GSM8K, and EMRQA-mSQuAD datasets.} 
    \label{fig: reconstruction-examples-appendix} 
    \end{figure*}

\clearpage

\begin{figure}[h]
    \centering
    \subfigure[AGNews/Before]{\includegraphics[width=0.235\textwidth]{Figures/semantic_similarity_ag_news_adamw.pdf}}
    \subfigure[AGNews/After]{\includegraphics[width=0.235\textwidth]{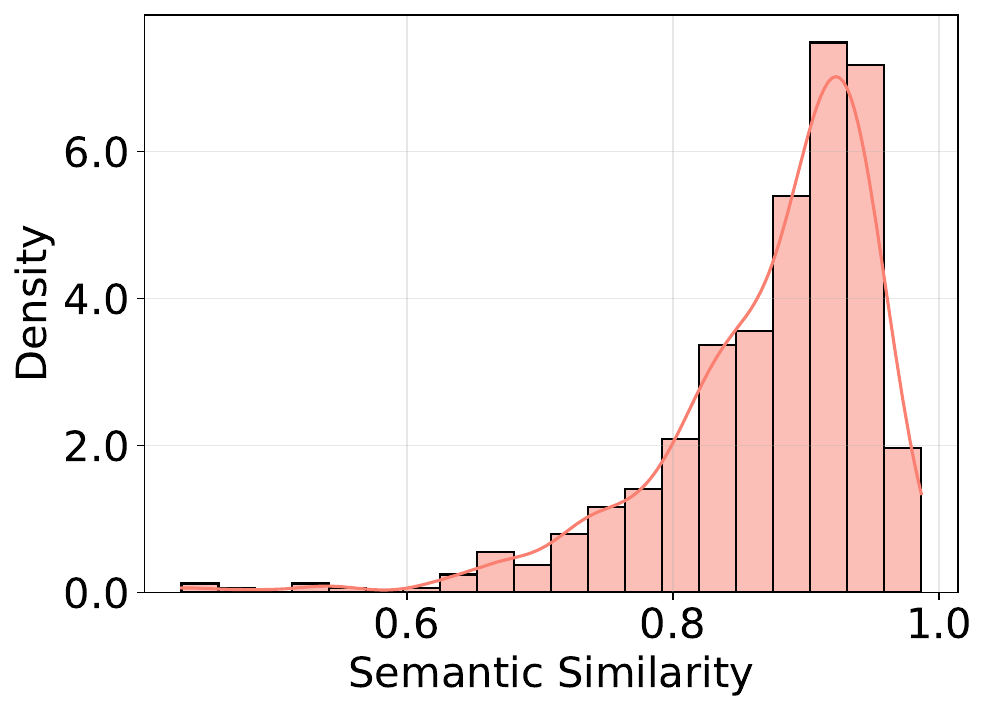}}
    \subfigure[SquAD/Before]{\includegraphics[width=0.235\textwidth]{Figures/semantic_similarity_squad_adamw.pdf}}
    \subfigure[SquAD/After]{\includegraphics[width=0.235\textwidth]{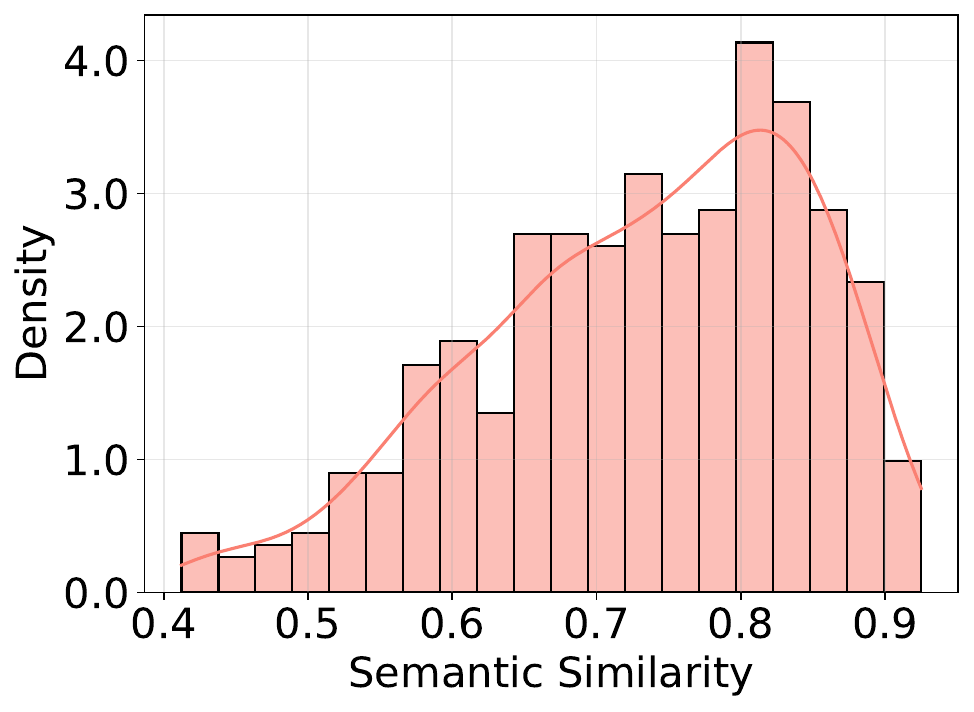}}
    \subfigure[mSQuAD/Before]{\includegraphics[width=0.235\textwidth]{Figures/semantic_similarity_emrqa_msquad_adamw.pdf}}
    \subfigure[mSQuAD/After]{\includegraphics[width=0.235\textwidth]{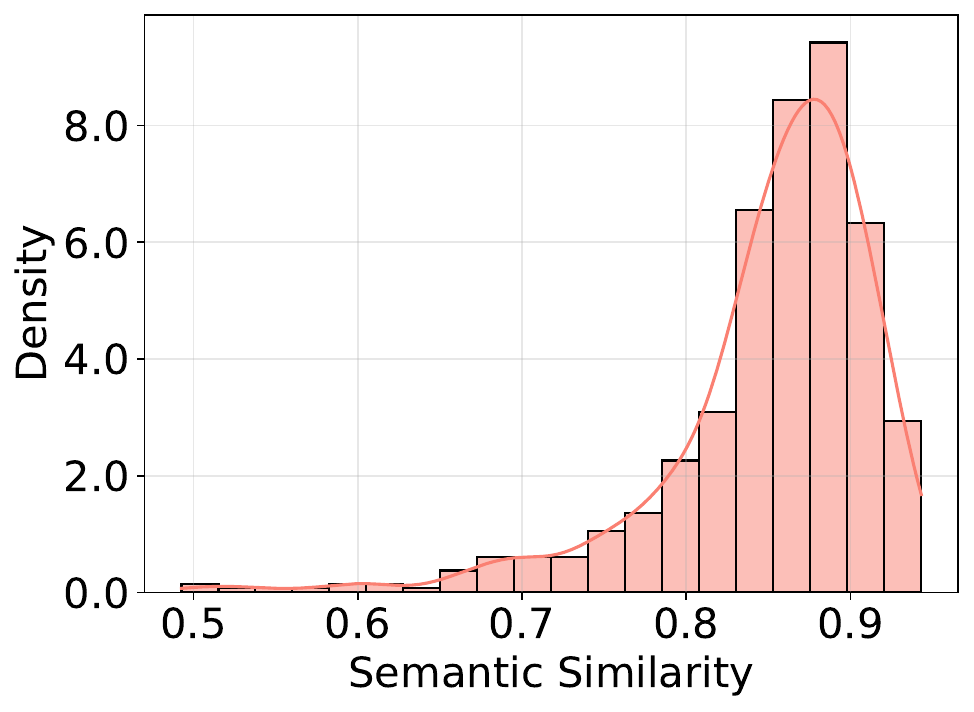}}
    \subfigure[GSM8K/Before]{\includegraphics[width=0.235\textwidth]{Figures/semantic_similarity_gsm8k_adamw.pdf}}
    \subfigure[GSM8K/After]{\includegraphics[width=0.235\textwidth]{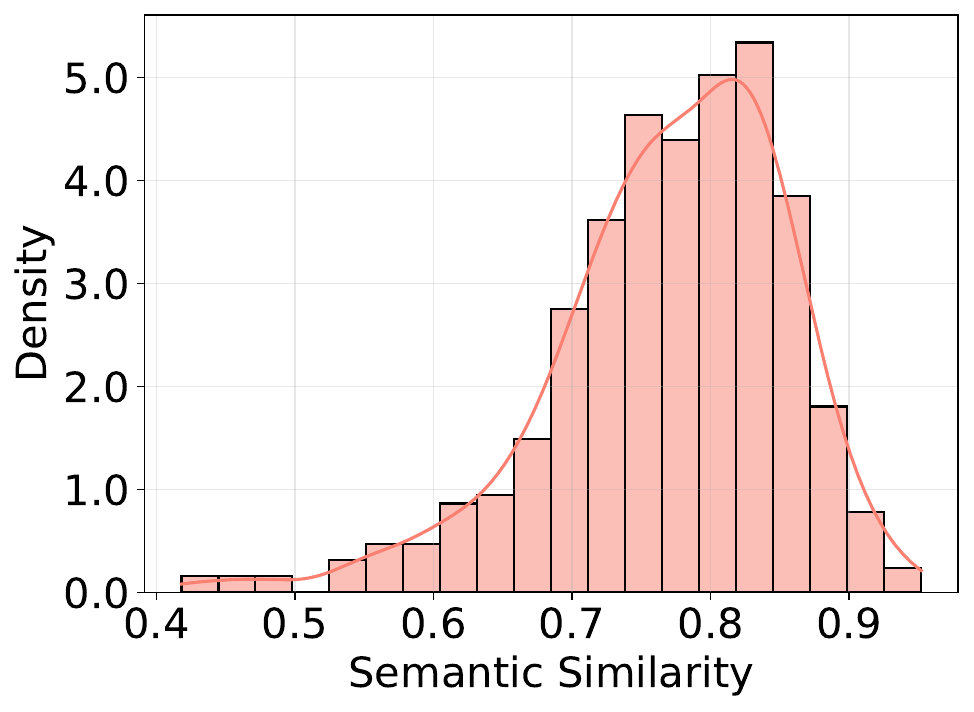}}
    \caption{The semantic similarity distribution of reconstructed samples before and after LLM refinement for Qwen2-1.5B. }
    \label{fig: LLM-Refinement} 
\end{figure}


\section*{Appendix E.~Attack Larger Models}
\label{Larger models}
We implemented \textit{\sysname} on larger models, including Qwen2-7B and Llama3-8B, under the same LoRA settings in Section 5.5. The goal is to evaluate whether \textit{\sysname} can be applied to more sophisticated language models. We illustrate the experiment results in Tab. \ref{tab: larger-models}. From the results, we observe that \textit{\sysname} maintains strong reconstruction performance on both larger models and achieves \textit{even better} performance under the AdamW optimizer than that of their smaller variants. This improvement can be attributed to the richer and more discriminative embedding representations in larger models, which enable more precise inversion from embeddings back to tokens. 

\begin{table}[t]
    \centering
    \caption{The reconstruction performance of \textit{\sysname} on larger language models.}
    \small
    \setlength{\tabcolsep}{4pt}
    \begin{tabular}{lcccc}
        \toprule
        \multirow{2}{*}{\textbf{Dataset}} &
        \multirow{2}{*}{\textbf{Optimizer}} &
        \multicolumn{1}{c}{\textbf{Qwen2-7B}} &
        \multicolumn{1}{c}{\textbf{Llama3-8B}} \\
        \cmidrule(r){3-4}
        & & Rate$|$Sim.$\pm$Std & Rate$|$Sim.$\pm$Std \\
        \midrule
        \multirow{2}{*}{AGNews}
            & SGD   & 0.748$|$0.998$\pm$0.030 & 0.723$|$1.000$\pm$0.000 \\
            & AdamW & 0.761$|$0.900$\pm$0.003 & 0.713$|$0.899$\pm$0.017 \\
        \midrule
        \multirow{2}{*}{SQuAD}
            & SGD   & 0.700$|$0.990$\pm$0.068 & 0.711$|$0.995$\pm$0.047 \\
            & AdamW & 0.726$|$0.892$\pm$0.062 & 0.761$|$0.898$\pm$0.024 \\
        \midrule
        \multirow{2}{*}{mSQuAD}
            & SGD   & 0.743$|$0.977$\pm$0.091 & 0.729$|$0.972$\pm$0.100 \\
            & AdamW & 0.696$|$0.891$\pm$0.055 & 0.700$|$0.887$\pm$0.069 \\
        \midrule
        \multirow{2}{*}{GSM8K}
            & SGD   & 0.700$|$0.974$\pm$0.113 & 0.735$|$0.966$\pm$0.130 \\
            & AdamW & 0.787$|$0.891$\pm$0.070 & 0.804$|$0.893$\pm$0.060 \\
        \bottomrule
    \end{tabular}
    \label{tab: larger-models}
\end{table}

\begin{table}[t]
    \centering
    \caption{LLM-refined reconstruction performance on GPT-2 and Qwen2-1.5B.}
    \small
    \setlength{\tabcolsep}{4pt}
    \begin{tabular}{lcccc}
        \toprule
        \multirow{2}{*}{\textbf{Dataset}} &
        \multirow{2}{*}{\textbf{Optimizer}} &
        \multicolumn{1}{c}{\textbf{GPT-2}} &
        \multicolumn{1}{c}{\textbf{Qwen2-1.5B}} \\
        \cmidrule(r){3-4}
        & & Rate$|$Sim.$\pm$Std & Rate$|$Sim.$\pm$Std \\
        \midrule
        \multirow{2}{*}{AGNews}
            & Before & 0.744$|$0.779$\pm$0.115 & 0.733$|$0.830$\pm$0.102 \\
            & After  & 0.745$|$0.788$\pm$0.110 & 0.734$|$0.873$\pm$0.081 \\
        \midrule
        \multirow{2}{*}{SQuAD}
            & Before & 0.734$|$0.682$\pm$0.119 & 0.531$|$0.668$\pm$0.112 \\
            & After  & 0.735$|$0.679$\pm$0.118 & 0.542$|$0.733$\pm$0.113 \\
        \midrule
        \multirow{2}{*}{mSQuAD}
            & Before & 0.739$|$0.806$\pm$0.078 & 0.736$|$0.767$\pm$0.077 \\
            & After  & 0.739$|$0.808$\pm$0.076 & 0.736$|$0.851$\pm$0.066 \\
        \midrule
        \multirow{2}{*}{GSM8K}
            & Before & 0.654$|$0.524$\pm$0.116 & 0.595$|$0.663$\pm$0.084 \\
            & After  & 0.635$|$0.522$\pm$0.108 & 0.596$|$0.770$\pm$0.085 \\
        \bottomrule
    \end{tabular}
    \label{tab: LLM-Refinement}
\end{table}

\section*{Appendix F.~LLM Refinement}

Under the AdamW optimizer, reconstructed samples preserve key keywords but are often disorganized and difficult to read. To mitigate this issue, we feed the reconstructed text under the AdamW optimizer into ChatGPT-5, and prompt it to refine and reorganize the samples into more coherent and readable sentences. We conduct this on the GPT-2 and Qwen2-1.5B models, as \textit{\sysname} achieves lower numerical performance on them. We compare the attack performance before and after refinement in Tab.~\ref{tab: LLM-Refinement}. The results show that LLM-based refinement improves the reconstruction performance on Qwen2-1.5B, but provides little to no benefit for GPT-2. Therefore, we regard this refinement as an optional auxiliary step that can help improve reconstruction quality when applicable. For better illustration, we provide the semantic similarity distribution of the reconstructed samples before and after the LLM refinement in Fig. \ref{fig: LLM-Refinement}.  


\end{document}